\documentclass[twocolumn]{aastex631}

\usepackage{amsmath,amssymb}
\usepackage{graphicx}
\usepackage{graphicx}
\usepackage{xcolor}
\usepackage[normalem]{ulem}
\usepackage{listings}
\usepackage{txfonts}

\newcommand{\unwise}{\textit{unWISE}}
\newcommand{\catwise}{\textit{catWISE}}
\newcommand{\healpix}{{\sl HEALPix}}

\def\deg{\ensuremath{^\circ}}
\providecommand{\gaia}{\textit{Gaia}}
\newcommand\gdrtwo{\gaia~DR2}
\newcommand\gdrthree{\gaia~DR3}

\newcommand{\muas}{\mu \rm{as}}

\begin{document}

\title{The Milky Way as Seen by Classical Cepheids I: Distances Based on Mid-infrared Photometry}
\shorttitle{Galactic classical Cepheid distances}
\shortauthors{Skowron et al.}

\correspondingauthor{Dorota M. Skowron}
\email{dszczyg@astrouw.edu.pl}

\author[0000-0001-9439-604X]{Dorota M. Skowron}
\affiliation{Astronomical Observatory, University of Warsaw, Al. Ujazdowskie 4, 00-478 Warsaw, Poland}

\author[0000-0002-1777-5502]{Ronald Drimmel}
\affiliation{INAF - Osservatorio Astrofisico di Torino, via Osservatorio 20, 10025 Pino Torinese (TO), Italy}

\author[0000-0002-2604-4277]{Shourya Khanna}
\affiliation{INAF - Osservatorio Astrofisico di Torino, via Osservatorio 20, 10025 Pino Torinese (TO), Italy}

\author[0000-0003-1732-2412]{Alessandro Spagna}
\affiliation{INAF - Osservatorio Astrofisico di Torino, via Osservatorio 20, 10025 Pino Torinese (TO), Italy}

\author[0000-0003-3793-8505]{Eloisa Poggio}
\affiliation{INAF - Osservatorio Astrofisico di Torino, via Osservatorio 20, 10025 Pino Torinese (TO), Italy}

\author[0000-0002-5080-7027]{Pau Ramos}
\affiliation{National Astronomical Observatory of Japan, Mitaka-shi, Tokyo 181-8588, Japan}

\begin{abstract}

Classical Cepheids  are the archetype of the standard candle, thanks to the period--luminosity relation which allows to measure their intrinsic brightness.  They are also relatively young and bright, potentially making them excellent tracers of the young stellar population that is responsible for shaping the visible aspect of our Galaxy, the Milky Way.
However, being observers embedded in the dusty interstellar medium of the Galaxy, deriving reliable photometric distances to classical Cepheids of the Milky Way is a challenge. The typical approach is to use “reddening-free” indices, such as Wesenheit magnitudes, to obviate the need for an extinction correction. However, this approach could lead to unknown systematics -- especially toward the inner Galaxy -- as its assumption of a universal total-to-selective extinction ratio is not satisfied, particularly in lines-of-sight where the extinction is high and crosses spiral arms.  We instead estimate new distances for 3424 Cepheids based on mid-IR photometry from WISE, which suffers minimally from extinction, and by adopting a 3D extinction map to calculate the necessary (albeit small) extinction corrections.
We show that our distances are consistent with \gaia’s parallaxes for the subset with relative parallax errors smaller than 10\%, verifying that our mean distance errors are of the order of 6\% and that the mean parallax zero point for this subsample is $7\: \muas$.

\end{abstract}


\section{Introduction}
\label{sec:intro}

Classical Cepheids are young massive and intermediate-mass stars pulsating in radial modes. They serve as distance indicators within the Milky Way and local galaxies, they are excellent tracers of the young stellar population, and they are successfully used to test stellar models. We refer the reader to \cite{Bono2024} for a detailed review of Cepheid properties and applications.

The number of known classical Cepheids in our Galaxy has significantly increased over the past five years with the discoveries from the Optical Gravitational Lensing Experiment (OGLE, \citealt{Udalski2018,Soszynski2020}), the Asteroid Terrestrial-impact Last Alert System (ATLAS, \citealt{Heinze2018}), the All Sky Automated Survey for Supernovae (ASAS-SN, \citealt{Jayasinghe2018,Jayasinghe2019}), the Zwicky Transient Factory (ZTF, \citealt{Chen2020}), and the \gaia{} mission \citep{Ripepi:2023}. The up-to-date list of Galactic classical Cepheids is maintained and regularly updated by \citet[][hereafter P21]{Pietrukowicz:2021} and currently contains 3659 stars.

These catalogs led to a great number of Milky Way studies, including that of the Galactic warp \citep{Skowron2019a,Skowron2019b,Dehnen2023,Cabrera-Gadea2024}, the spiral structure \citep{Minniti2021,Lemasle2022,Bobylev2022,GaiaCollaboration2023,Semczuk2023}, the rotation curve and the outer disk \citep{Mroz2019,Ablimit2020,Drimmel:2023resonance,Bobylev2023,Beordo2024}, just to name a few.
All such studies require calculating distances to classical Cepheids, and in each case this is done with the use of period--luminosity (PL) relations \citep{Leavitt1908}. However, various studies have different approaches to the exact method of distance estimation. For example \cite{Skowron2019a,Skowron2019b}, hereafter S19, used mid-infrared (mid-IR) data from the Wide-field Infrared Survey Explorer (WISE, \citealt{Wright2010}) and {\it Spitzer} Space Telescope \citep{Werner2004}, together with mid-IR PL relations and an extinction correction; \cite{GaiaCollaboration2023} used photometric data from the third \gaia\ data release (hereafter \gdrthree) and period--Wesenheit--metallicity relations calibrated for the {\em Gaia} bands to avoid the extinction corrections; while \cite{Lemasle2022} use mid-IR data from WISE and period--Wesenheit relations calibrated for mid-IR WISE bands, also without an extinction correction. 

\begin{figure}
    \centering
    \includegraphics[width=0.99\columnwidth]{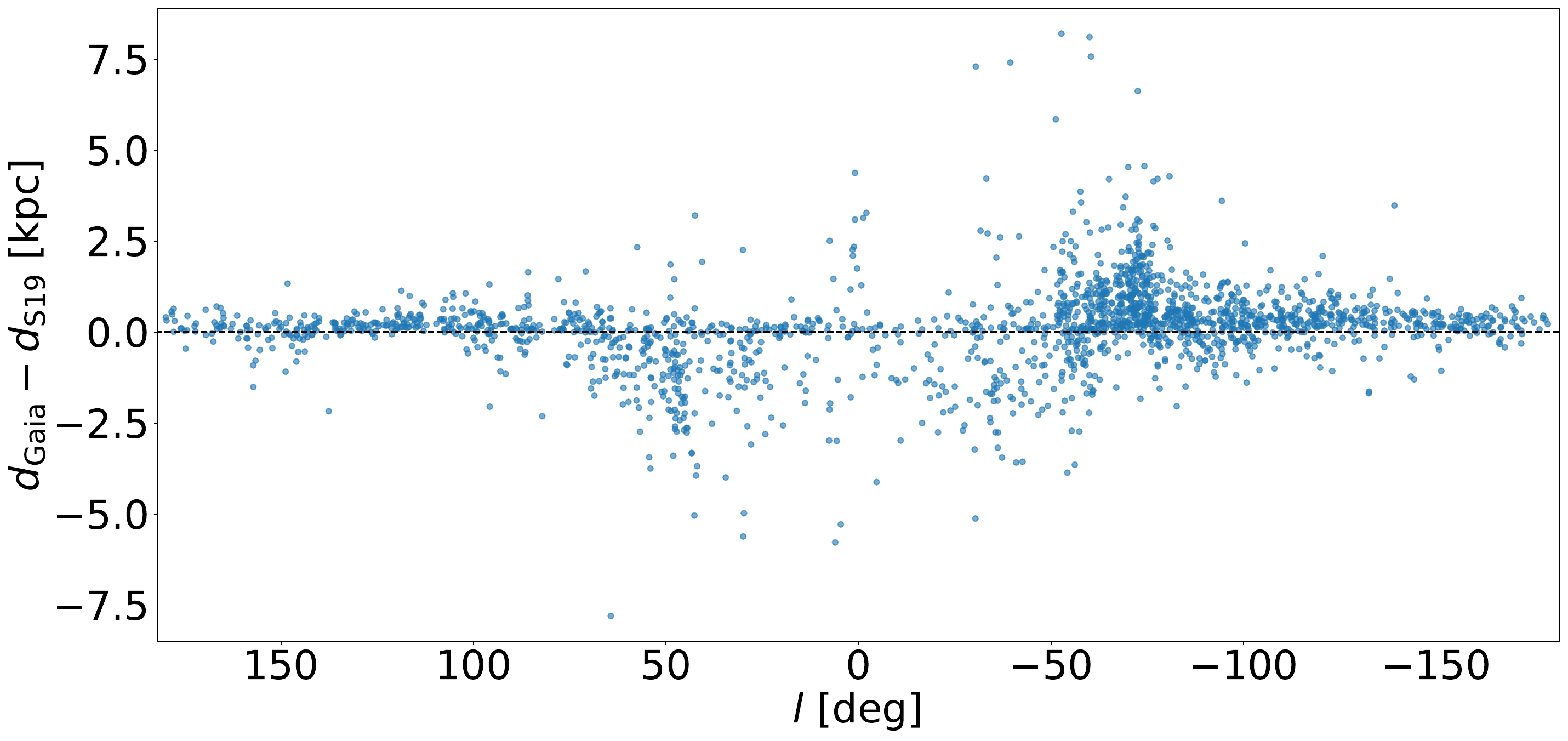}
    \caption{The difference in Cepheid distances calculated with different methods, plotted against Galactic longitude, where $d_{\rm Gaia}$ is from \cite{GaiaCollaboration2023} and $d_{\rm S19}$ from S19. The Galactic center is at $l=0^\circ$, while the anticenter at $l=\pm180^\circ$. The mean general offset of about $200$~pc is calculated in the range $l>120^\circ$ and $l<-120^\circ$. The distance difference for $-100^\circ<l<100^\circ$ has a mean absolute value of about $930$~pc and a dispersion of $1400$~pc.}
    \label{fig:distance_comparison_intro}
\end{figure}

Figure~\ref{fig:distance_comparison_intro} shows the comparison of distances between the study of \cite{GaiaCollaboration2023} and S19, where the distance difference is plotted against the Galactic longitude, with $l=0^\circ$ pointing toward the Galactic center. It is striking how large the distance differences are (despite the reported distance uncertainties of about $5-10\%$), and calls into question the detailed studies of Galactic structure with classical Cepheids. A similar comparison between distances from studies of S19 and \cite{Lemasle2022} shows discrepancies of the same order of magnitude, though not necessarily the same sign (B. Lemasle, private communication).

Interestingly, the differences in distance (Figure~\ref{fig:distance_comparison_intro}) are very large in the direction of the Galactic center, but negligible toward the anticenter ($l=\pm180^\circ$). The mean general offset of about $200$~pc between the two datasets is expected and results from the different approach to distance calculation. Instead, the distance differences in sight lines toward the Galactic center ($-100^\circ<l<100^\circ$) have a mean absolute value of about $930$~pc and a dispersion of $1400$~pc. 
This suggests that the main factor contributing to distance uncertainty is tied to extinction, which is highest toward the Galactic center. Indeed, the usage of the Wesenheit index in the period-Wesenheit relation to avoid the extinction correction has an underlying assumption that the ratio of total-to-selective extinction is constant throughout the Galaxy (i.e. a universal extinction curve), which is not true in regions of the disk with high levels of extinction, where the extinction curve is known to be highly variable \citep{Fitzpatrick2007,  Schlafly2016, Nataf2016, Maiz-Apellaniz2018, Fouesneau2022, Zhang2023, Zhang2025}.
Since a significant number of classical Cepheids are located in such regions, it is preferable to use distance estimates less sensitive to the assumption of a universal extinction curve. 

On the other hand, the use of the PL relation instead of the period--Wesenheit index relation requires accounting for interstellar extinction. This is typically handled by moving to the mid-IR wavelengths, where the extinction is $\sim$10 times smaller than in the optical. However, even though mid-IR extinction values are generally very low, they are non-negligible within the Galactic plane and toward the inner disk, where a large fraction of the Galaxy's Cepheids are located. 
It is therefore necessary to use three-dimensional extinction maps to account for extinction. Motivated by Figure~\ref{fig:distance_comparison_intro}, we decided to derive a new set of accurate distances based on the mid-IR data, mid-IR PL relations and mid-IR extinction corrections from 3D extinction maps, with the goal of producing a catalog of Cepheid distances that can be used for detailed Galactic structure studies.

The paper is organized as follows: In section \ref{sec:data} we describe the compilation of our source data, in section \ref{sec:results} the calculation of the new distances, their validation and a comparison with previously published distances.  In the final two sections we discuss and summarize our results.

\section{Data}
\label{sec:data}

\subsection{Classical Cepheid Catalog}

We use the list of P21\footnote{\url{https://www.astrouw.edu.pl/ogle/ogle4/OCVS/allGalCep.listID}} as a source of classical Cepheid coordinates, periods and modes of pulsation. We used the version of the table from September 2023, already updated with classical Cepheids from \gdrthree.
The list originated from the compilation of 745 Cepheids provided in the General Catalog of Variable Stars (GCVS; \citealt{Samus2017}) and now includes discoveries from various surveys, with the major contribution coming from OGLE (1690 stars), ZTF (422), \gaia\ (298), ASAS-SN (166), ATLAS (139) and ASAS (109). Because some surveys use automated variable star classification pipelines, their Cepheid catalogs may contain misclassifications.
For this reason, each classical Cepheid that is on the list of P21 has been visually confirmed by the authors to ensure the sample purity. This means that not all stars classified as Cepheids by the above surveys have been included in the list. 
\begin{figure*}
    \centering
    \includegraphics[width=1.99\columnwidth]{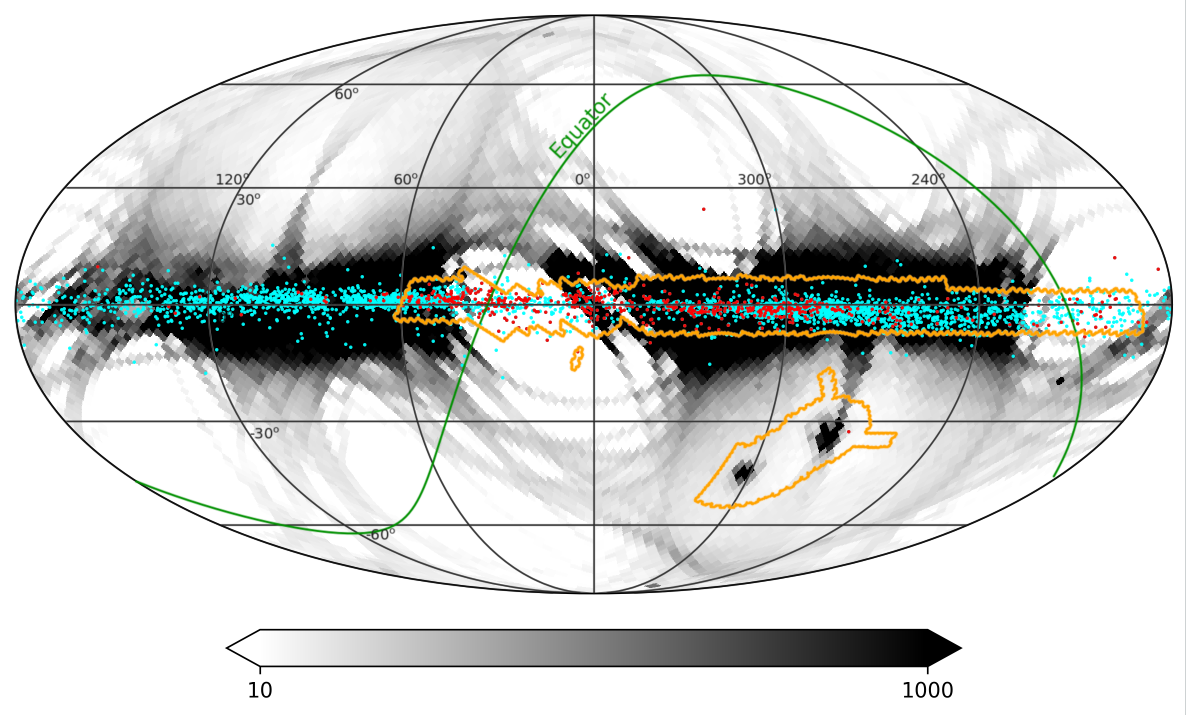}
    \caption{The completeness of the classical Cepheid sample in Galactic coordinates. The OGLE footprint is shown with orange. The ZTF sky coverage extends north of the Equator (green line). The \gaia, ASAS-SN and ATLAS surveys observe the entire sky. The gray background shows the distribution of 5.5 million \gdrthree{} variable stars (excluding AGN \& planetary transits) with \textit{num\_selected\_g\_fov} $>40$ at \textit{HEALPix} level 5.
    The cyan points show the location of our entire Milky Way Cepheid catalog (3659 sources), and in red we overplot only those that are exclusively detected outside of \gaia~(605 sources), using the flag \textit{in\_vari\_cepheid}=False.}
    \label{fig:ceph_selfunc_sky}
\end{figure*}

The completeness of the Cepheid catalog is difficult to assess, as the contributing surveys have different sky coverage, time-span of observations, number of data-points per star, magnitude range, and data processing pipelines. 
Figure~\ref{fig:ceph_selfunc_sky} shows the sky areas observed by the most contributing surveys. The OGLE footprint is marked with orange polygons, and the ZTF range spans north of the Celestial Equator (in green). The remaining surveys (\gaia, ASAS-SN, ATLAS) are not marked, as they observe the entire sky. All 3659 Galactic classical Cepheids from the list of P21 are overplotted with cyan and red points. 
The completeness of the classical Cepheid catalog within the OGLE footprint is estimated to be $90\%$ down to $I\simeq19.5$~mag \citep{Udalski2018,Soszynski2020}, and from ZTF it is $63.3\%$ down to $r\simeq20.6$~mag \citep{Chen2020}. The completeness of the ASAS-SN and ATLAS catalogs is not quantified, but depends on the brightness, amplitude of variability (being lower for smaller amplitudes), and crowding effects, which play a significant role due to the larger pixel size \citep{Jayasinghe2018}. 

\begin{figure}
    \centering
    \includegraphics[width=1.03\columnwidth]{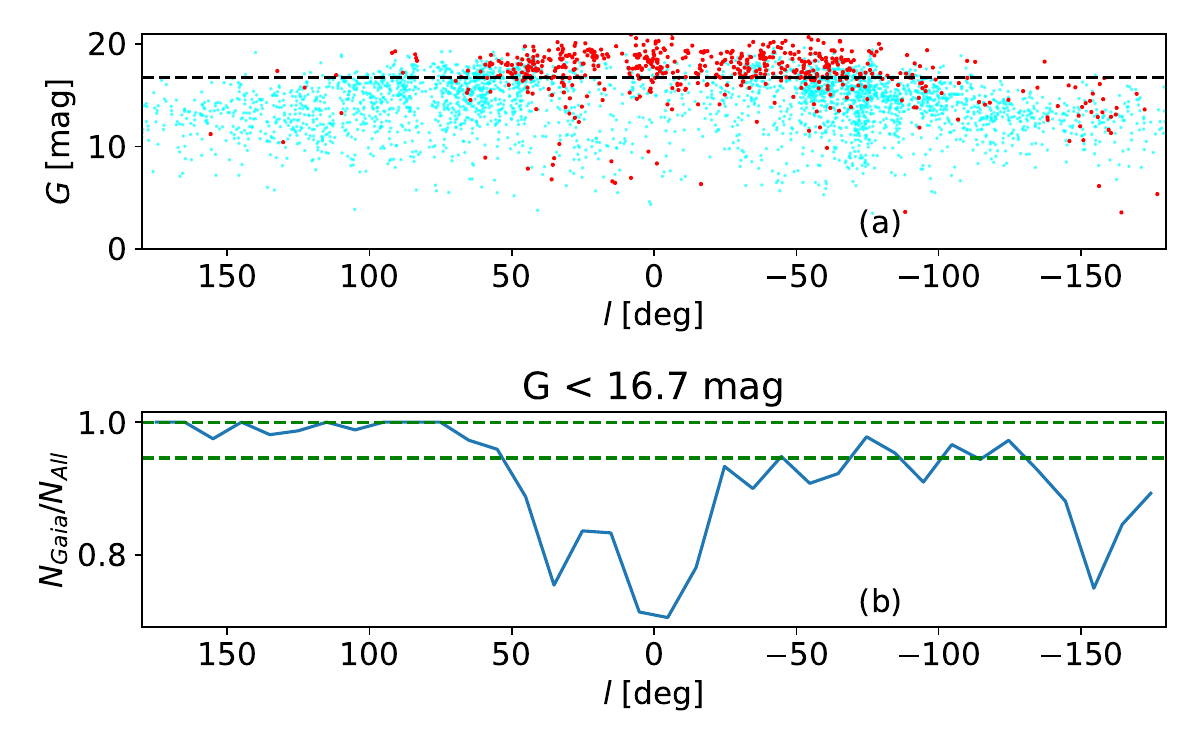}\caption{Completeness estimate of \gaia~ Cepheids across Galactic longitude ($l$) in 10$^\circ$ wide bins. Panel (a) shows the distribution of $l$ as a function of $G$ magnitude. In cyan, are all Cepheids detected with the \gaia\ survey, while in red, are all exclusively discovered outside of \gaia. The black dotted line indicates the magnitude limit ($84^{th}$ percentile) of \gaia\ Cepheids ($G < 16.7$ mag). Panel (b) shows the contribution of Cepheids from \gaia\ ($N_{\rm Gaia}$) to the entire catalog ($N_{all}$) as a function of $l$, restricting the sample to within the magnitude limit of \gaia, where this metric provides a measure for the completeness of Cepheids observed by \gaia. The two green dashed lines indicate completeness of 0.95 and 1.0.}
    \label{fig:ceph_selfunc_sky2}
\end{figure}

In the case of \gaia, the completeness of the Cepheid catalog depends on the efficiency of the classification, resulting in a Selection Function that strongly depends on the \gaia\ scanning law. Figure~\ref{fig:ceph_selfunc_sky} shows in gray scale the sky distribution of the variable sources in \gdrthree\ \citep{Eyer:2023}, excluding AGN \& planetary transits, whose number of epoch observations (\textit{num\_selected\_g\_fov}) is greater than 40. 
The inhomogeneous sky coverage is apparent. Among all the 3659 Cepheids in our catalog, there are 605 stars not classified as Cepheids in \gaia\ (\textit{in\_vari\_cepheid = False}) shown in red. The majority of unidentified Cepheids in \gaia\ are fainter than $G\geq17$~mag, and found toward the inner Galaxy. Nevertheless, even at brighter magnitudes we see a higher level of incompleteness at Galactic longitudes where the number of epoch observations fall below 40 
(See Figure \ref{fig:ceph_selfunc_sky2}). We can expect a much more efficient detection and classification of Cepheids in the next \gaia\ data release expected to cover 66 months of observations. 

In summary, thanks to the combined coverage of the different surveys, we believe that our sample is highly complete to at least $G = 18$~mag ($I=19.5$~mag) within the footprint of OGLE, though will be smaller in highly extincted areas toward the Galactic center. Outside this footprint our completeness will be somewhat lower. Thanks to the intrinsic brightness of Cepheids and limited extinction, we are likely seeing all Milky Way Cepheids in the second and third quadrant of the Galaxy ($90\deg < l < 270\deg$), and most Cepheids outside the Galactic bulge region (outside $30\deg$ of the Galactic center) in the first and fourth quadrants, with extinction more or less limiting our view in selective directions.

\subsection{Mid-IR Photometry}
\label{sec:photometry}

We use the WISE catalogs as a source of mid-IR photometry for our sample of classical Cepheids. The WISE space telescope has observed the whole sky in  $W1$ [3.4$\,\mu$m], $W2$ [4.6$\,\mu$m], $W3$ [12$\,\mu$m], and $W4$ [22$\,\mu$m] bands during 2009-2011, and in $W1$ and $W2$ since 2013, after the coolant depletion, as the Near-Earth Object Wide-field Infrared Survey Explorer (NEOWISE, \citealt{Maizner2011}). The WISE data release -- the AllWISE Source Catalog \citep{Cutri2013}\footnote{\href{http://wise2.ipac.caltech.edu/docs/release/allwise/expsup/index.html}{WISE data release notes}} provides coordinates and flux measurements for almost 750 million sources in $W1$, $W2$, $W3$, and $W4$ bands. 
The unWISE Catalog \citep{Schlafly2019} is based on the co-added WISE images in $W1$ and $W2$ from both the WISE and NEOWISE phases of the mission and therefore is much deeper than the original AllWISE Catalog, with about five times larger total exposure time. In addition, an improved modeling of crowded regions was implemented to simultaneously determine fluxes and positions of all sources, providing more accurate photometry with respect to AllWISE.
The CatWISE Catalog \citep{Marocco2021} is also based on coadded WISE and NEOWISE images in $W1$ and $W2$, but uses the AllWISE software to measure the source brightness while taking into account possible proper motion over the observational baseline.

We cross-matched our list of classical Cepheids with the above catalogs using a 1" search radius. 
For Cepheids brighter than $W1 < 8.25$~mag, the AllWISE $W1$ magnitudes are generally considered more reliable than either CatWISE or unWISE, as the photometric pipelines for these two catalogs were optimized for fainter magnitudes, and so could be significantly biased for $W1 < 8$~mag where WISE images begin to be saturated. Also, when cross-matching our sample of Cepheids to these catalogs we noticed that some bright Cepheids were missing from both CatWISE and unWISE, confirming the need to include AllWISE data, at least at the bright end. Furthermore, at fainter magnitudes we found that unWISE was more complete than CatWISE for our Cepheid sample, so we adopt the unWISE photometry over that of CatWISE.

Moreover, for 534 bright Cepheids, we use data from the WISE All-Sky Release Source Catalog\footnote{\href{https://wise2.ipac.caltech.edu/docs/release/allsky/expsup/sec2\_2.html}{WISE All-Sky release notes}} in place of the AllWISE magnitudes,
to avoid the problematic AllWISE measurements that were made during the Post-Cryo mission phase, as suggested by the AllWISE team\footnote{ 
\href{https://wise2.ipac.caltech.edu/docs/release/allwise/expsup/sec2\_2.html\#cat\_phot}{AllWISE cautionary notes}}.
These are Cepheids brighter than $W1=$ 8.25~mag in AllWISE found in the ecliptic longitude ranges $89.4\deg<\lambda<221.8\deg$ and $280.6\deg<\lambda<48.1\deg$.

\begin{figure}
    \centering
    \includegraphics[width=0.49\textwidth]{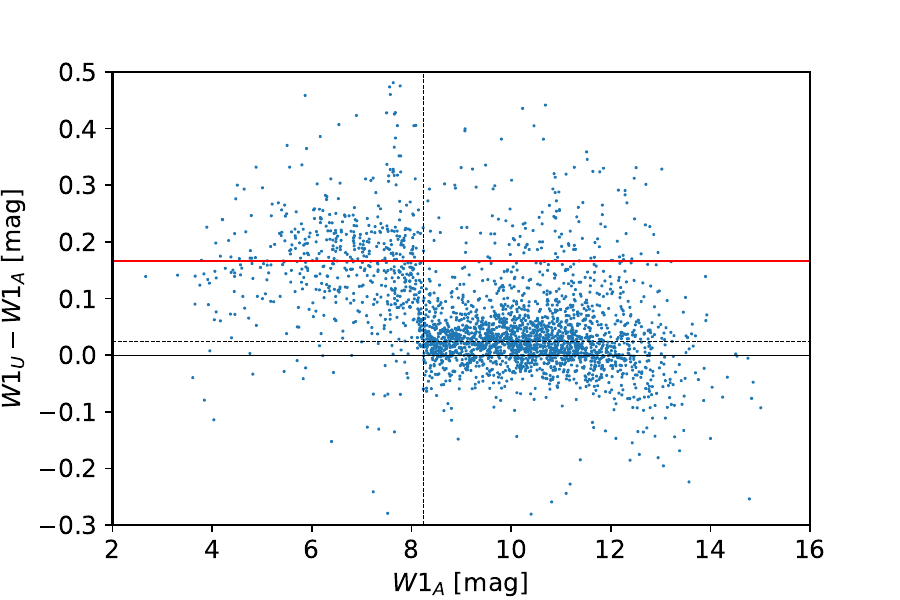}
    \caption{Difference in $W1$ magnitudes between AllWISE and unWISE as a function of $W1$ from AllWISE. The dashed horizontal line shows the mean offset between the two catalogs for $W1 > 8.25$ mag (0.024 mag), while the red horizontal line is the mean offset for magnitudes brighter than 8.25 mag (0.166mag).}
    \label{fig:diffW1}
\end{figure}

Since we wish to use the photometry from two different catalogs (unWISE and AllWISE), it is important to confirm that the two catalogs are on the same photometric system,  especially as we will use the $W1$ PL relation (Leavitt law) from \citet{Wang2018}, which was calibrated using AllWISE data. 
\citet{Schlafly2019} recommend applying a $\Delta W1 = W1_U - W1_A$ correction of 0.004 mag between the AllWISE ($W1_A$) and unWISE ($W1_U$) magnitudes, but also noted a clear magnitude dependence between the two magnitude systems (see their Figure~7). For our sample, we find an offset $\Delta W1$ of 0.024 mag for $W1_U > 8.25$~mag  (see Figure~\ref{fig:diffW1}), with little evidence of a magnitude term. Therefore we adopt 0.024 mag as the offset at these magnitudes to bring the unWISE photometry to the AllWISE photometric system before calculating distances.
 
\begin{figure}
    \centering
    \includegraphics[width=0.49\textwidth]{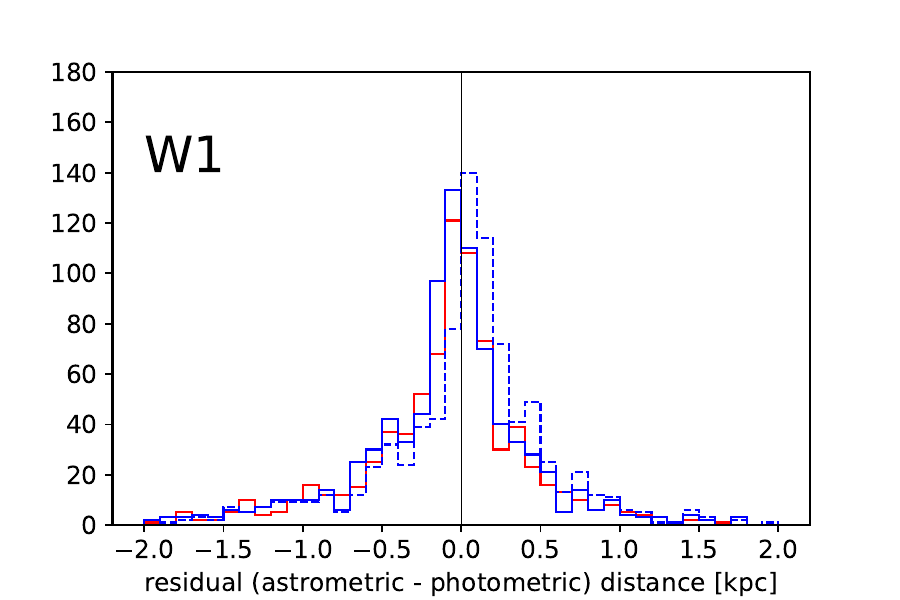}
    \includegraphics[width=0.49\textwidth]{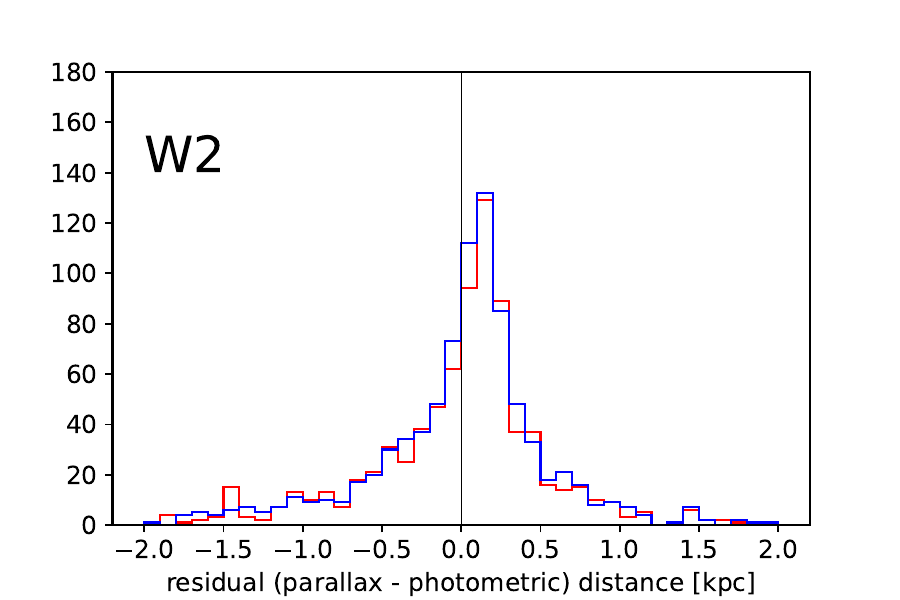}
    \caption{Distribution of the differences in the astrometric and photometric distances for bright ($W1_A < 8.25$~mag) Cepheids with good parallaxes using $W1$ (top panel) and $W2$ (bottom panel). The red histogram shows differences derived from the unWISE magnitudes and the blue from the AllWISE magnitudes.
    An offset of 0.024 mag has been applied to the unWISE magnitudes, while the differences between the astrometric and photometric distances derived from AllWISE $W1_A$ magnitudes are shown both with (solid line) and without (dashed line) a 0.142 offset applied to the AllWISE magnitudes for stars brighter than $W1 < 8.25$.}
    \label{fig:ddist_plx_phot}
\end{figure}

Meanwhile for magnitudes $W1_A < 8.25$~mag we find a significantly larger $\Delta W1$ of 0.166 mag (see Figure~\ref{fig:diffW1}). One might be tempted to interpret this as evidence that indeed the unWISE photometry should not be used at bright magnitudes, and to use only the AllWISE magnitudes.  However, a more careful comparison of the photometric distances derived from AllWISE and unWISE $W1$ magnitudes revealed that, once the photometric offsets between the two catalogs are taken into account, the unWISE $W1$ distances are of similar quality to those of AllWISE, even at bright magnitudes (See section \ref{sec:distances}). This is perhaps to be expected since saturation will be problematic for both catalogs. Moreover, judging from the comparison to the astrometric distances based on parallax, $(\varpi + 0.017~\muas)^{-1}$, for the subset of 520 bright ($W1_A < 8.25$~mag) Cepheids with good parallaxes ($\sigma_\varpi/\varpi < 0.1$ and $ruwe < 1.4$), it seems that the large $\Delta W1 = 0.166$~mag offset between the two catalogs at $W1 < 8.25$~mag is largely due to a bias in AllWISE (see upper panel in Figure~\ref{fig:ddist_plx_phot}). That is, the AllWISE $W1_A$ magnitudes at $W1_A < 8.25$~mag are biased with respect to fainter stars in AllWISE, and it is only after adding an additional $0.142$ $(=0.166-0.024)$ mag to these bright stars that the AllWISE photometric distances can be made consistent with the parallax-based distances at bright magnitudes. 

In contrast, we found that the $W2$ photometry (and distances) from AllWISE and unWISE seem to be quite consistent with each other, after adjusting for an offset of 0.024 mag, at both bright and faint magnitudes. However, the photometric distances using either set of $W2$ magnitudes, while consistent with each other, both show significant bias with respect to the astrometric distances from \gaia\ (see lower panel in Figure \ref{fig:ddist_plx_phot}). This bias is likely due to an unknown offset in the $W2$ magnitudes present in both catalogs at bright magnitudes.
Moreover, it has been shown that the Cepheid brightness in $W2$ (4.5 $\mu {\rm m}$) is affected by carbon monoxide absorption \citep{Marengo2010,Scowcroft2011,Monson2012} and so its use as a distance indicator has been questioned.
For these reasons we have decided to proceed using only the $W1$ magnitudes. 

\subsubsection{Crowding}
\label{sec:crowding}

Since we wish our sample to be as complete as possible, and we found that reliable distances could be derived even for the bright problematic sources, we make no selections based on either the unWISE or AllWISE flags. However, as our sample is primarily found in the Galactic plane, and given that the PSF of WISE $W1$ is large (6.1")\footnote{\url{https://wise2.ipac.caltech.edu/docs/release/allsky/}}, it is possible that in the most crowded regions of the Galaxy
the WISE photometry is potentially affected by crowding.  A useful measure of potential crowding effects provided by unWISE is the quality indicator {\em fracflux}, which is the PSF-weighted fraction of flux attributed to the source at its location.  A careful quantitative study of the sensitivity of {\em fracflux} to nearby sources was recently made by \citet{Saydjari2023} via experiments with injected sources, who use a similar (but improved) pipeline for the production of DECaPS2\footnote{Dark Energy Camera Plane Survey data release 2}. For their images they find little or no bias in the measured photometry for {\em fracflux} $> 0.5$, while the measured uncertainties are increasingly underestimated with decreasing {\em fracflux} (See lower-right panels of their Figures 21 and 22).  However, since these results are based on experiments with DECaPS imaging, it would be better to explore the quality of the photometric uncertainties with respect to {\em fracflux} with unWISE data. We check for a possible contribution from crowding to the unWISE magnitude errors by cross-matching unWISE and \gaia\ over a large area of the sky, and comparing the unWISE $W1$ magnitudes to $W1$ magnitudes derived from \gaia\ photometry, whose much higher resolution will be much less affected by crowding effects. We find no significant increase in the dispersion of these magnitude differences as a function of {\em fracflux}. See the discussion in Appendix \ref{Appendix:A} for further details. 

\begin{figure}
    \centering  \includegraphics[width=0.49\textwidth]{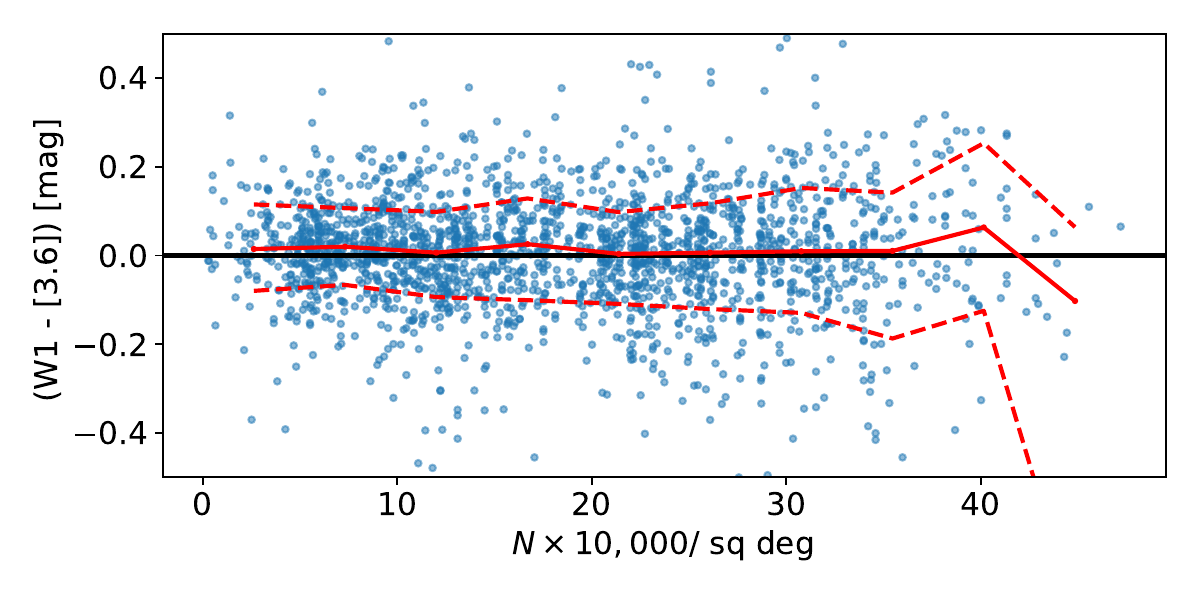}
    \caption{The difference between the WISE $W1$ (averaged) and \textit{Spitzer} [3.6] $\mu {\rm m}$ magnitudes for the Cepheids in this study, shown as a function of source density per square degree (blue points). The running median and the 16$^{th}$ and 84$^{th}$ percentiles are shown as red solid and dashed lines, respectively.}
    \label{fig:density_phot}
\end{figure}

However, the above analysis was necessarily conducted over an area of the sky with little or no extinction, outside the Galactic plane, where crowding is minimal. Therefore it only tests the effect on the photometry when there is a nearby, possibly blended source, but does not test for the additional biases that may come from a noisier background present in a crowded field. 
In order to check this, we compare our $W1$ photometry with \textit{Spitzer} [3.6] photometry as these passbands are quite similar, while \textit{Spitzer} has a much smaller PSF of 1.66" \citep{Fazio:2004}. Specifically, we cross-match within a one arcsecond search radius our Cepheids with the \textit{Galactic Legacy Infrared Midplane Survey Extraordinaire (GLIMPSE)} catalogue, including the extensions of their survey: \textit{GLIMPSE II, GLIMPSE 3D, GLIMPSE 360, Deep GLIMPSE, Vela-Carina, GLIMPSE Proper, APOGLIMPSE, "Spitzer Mapping of the Outer Galaxy" (SMOG) and "A Spitzer Legacy Survey of the Cygnus-X Complex" (Cygnus-X)} programs \citep{glimpse1_Benjamin:2003,Churchwell:2009}. 
We recover [3.6] magnitudes for 2019 Cepheids within the Glimpse footprint.
The $W1$ and [3.6] passbands are quite similar, so we perform a straight difference, ($W1-$[3.6]). We find that the median difference is 0.012 magnitudes (RMS = 0.17), which decreases to $-0.0015$ magnitudes (RMS  = 0.13) if we restrict to sources fainter than [3.6] $>$ 8 mag. In general we do see higher ($W1-$[3.6]) differences (i.e. scatter) toward the inner Galaxy (RMS = 0.25 
for Cepheids within $50^\circ$ of the Galactic center), where crowding can be an issue. To directly test the effect of crowding on the photometry, we tile the \textit{Glimpse} compilation using the \healpix{} pixelation scheme \citep{Gorski:2005} at level 6 (Nside = 64), resulting in an area per pixel of 0.84 square degrees. 
We counted the number of \textit{Glimpse} sources within each HEALPix to derive the source density per square degree at the $W1$/[3.6] wavelengths. The ($W1-$[3.6]) difference with respect to the number density in each HEALPix, shown in Figure~\ref{fig:density_phot}, confirms that there is a slight increase in the RMS with respect to the source density of the field, but there is no significant bias.
It is also worth noting, that some of the above scatter is a result of the stellar variability. The mean brightness of a Cepheid is based on about 10 epochs in the case of WISE, while typically on 2 epochs in the case of \textit{Spitzer}. This means that the uncertainty of estimating a mean Cepheid magnitude from \textit{Spitzer} is over twice the uncertainty from WISE (for details see Section~\ref{sec:uncertainties}).
We therefore conclude that crowding has no substantial effect on the $W1$ magnitudes, and for simplicity decided not to include any additional error terms.

\subsubsection{Photometric uncertainties}
\label{sec:uncertainties}

\begin{figure}
    \centering
    \includegraphics[width=1.\columnwidth]{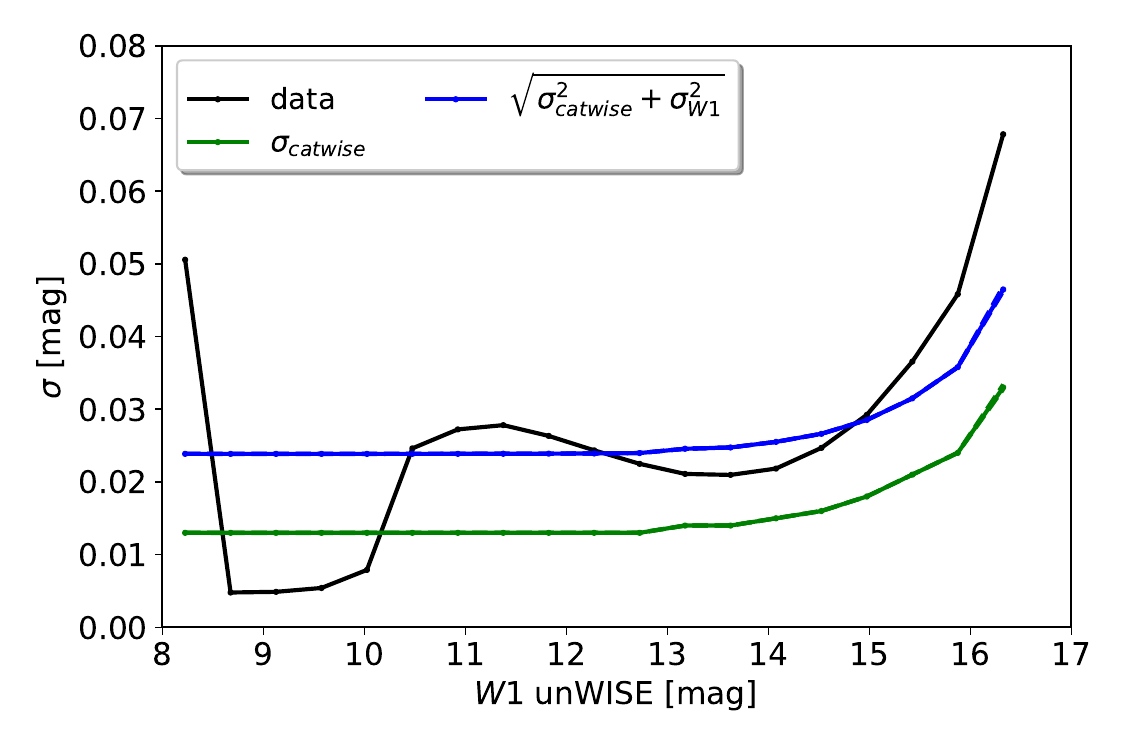}
    \caption{Standard deviation of the difference in $W1$ photometry between \unwise{} and \catwise{} shown in black, as a function of $W1$ from \unwise{}, for a field far away from the Galactic plane ( $205^{\circ} <l<225^{\circ}$ \& $10^{\circ}<b<20^{\circ}$) and $G<17$ mag. The green curve is the nominal uncertainty in \catwise{} $W1$, while in blue we boost this value by adding in quadrature additional uncertainty $\sigma_{W1}$ from Eq.~\ref{eq:sigmaW1}, in order to scale up to the observed curve in black.}
    \label{fig:unwise_catwise_errors}
\end{figure}

For the photometric uncertainties $\sigma_{W1}$ we take the AllWISE and WISE All-Sky uncertainties as reported, while for the unWISE catalog, that only reports a formal error on the fluxes, we adopt the following error model for the magnitude uncertainty: 
\begin{equation} \label{eq:sigmaW1}
    \sigma_{W1} = \sqrt{(dflux/flux)^2 + 0.02^2},
\end{equation}
where the $flux$ and $dflux$ are $W1$ flux and formal flux uncertainties as given in unWISE. The 0.02 in the second term defines a floor in the uncertainty of the unWISE $W1$ magnitudes due to unaccounted systematics.  We arrive at this value using the differences between the unWISE and CatWISE $W1$ magnitudes for a large sample of sources, and estimating the additional uncertainty needed in unWISE to account for the observed standard deviation $(\sigma)$ of these differences for unWISE magnitudes $W1 > 8.5$~mag, on the assumption that the CatWISE uncertainties are correct.  
An additional Gaussian dispersion of 0.02 mag is needed to account for the observed $\sigma$ of the differences at magnitudes fainter than about $W1_U\simeq 10.5$~mag (see Figure~\ref{fig:unwise_catwise_errors}), which we attribute to unWISE uncertainties.
This gives perhaps a conservative estimate on the unWISE magnitude errors but, as we show in the following section, the  uncertainties of our photometric distance moduli derived from the PL relation are still almost an order of magnitude larger than these estimated photometric errors. 

Another source of uncertainty arises when estimating the mean brightness of Cepheids, due to their variability. That is, determining the mean brightness from a few random measurements taken over time introduces a larger uncertainty for a variable star than for a constant one, and the greater the amplitude of variability, the greater the uncertainty. The unWISE and AllWISE source catalogs provide single brightness measurements performed on coadded frames, so no information about the amplitude of variability is contained in the reported photometric uncertainty. The mid-IR light curves for some Cepheids in our sample are available from the AllWISE Multiepoch Photometry Database, but they are often grouped around one or two epochs, which is equivalent to one or two light curve data points, given the fairly long pulsation periods of Cepheids. However, for some Cepheid light curves the number of epochs is sufficient to determine their mid-IR amplitudes and these have been used by \cite{Skowron2019a} to estimate the conversion factor between the mid-IR and optical amplitudes, as the latter are available for all Cepheids in the sample, contrary to mid-IR amplitudes. They found that the rms scatter in all mid-IR bands is half the rms scatter in the {\it I}-band and $\sim$30\% the rms scatter in the {\it V}-band (i.e., ${Amp}_{W1} = 0.5 \: {Amp}_{I}$ and ${Amp}_{W1} = 0.3 \: {Amp}_{V}$).
We use the {\it I}-band amplitude values provided by the OGLE survey, which are reliable, as all have been verified by eye, to estimate $W1$ amplitude values. However, not all Cepheids in our sample have the optical amplitudes from OGLE, therefore for the ones that do not, we use {\tt peak\_to\_peak\_bp} amplitudes from the {\tt gaiadr3.vari\_cepheid} table. We found that the conversion factor between {\it Gaia} BP band and WISE $W1$ is $0.32$ such that ${Amp}_{W1} = 0.32 \: {Amp}_{BP}$.

While doing the comparison, we noticed that some faint stars have unrealistically large BP amplitudes. The largest known amplitude of a classical Cepheid in the {\it I}-band is 1.38~mag (based on over $11,700$ classical Cepheids from OGLE; \citealt{Soszynski2019,Soszynski2020}), which translates to approximately $2.2$~mag in the BP band. Therefore, for all {\it Gaia} Cepheids with very large amplitudes, we adopt ${Amp}_{\rm BP} = 2.2$~mag. 

The mid-IR light curves of Cepheids are typically sinusoidal. The uncertainty of estimating the mean value for a sinusoidal from one random epoch is equal to $0.7$ of its amplitude (i.e. $Amp_{sin}/\sqrt{2}$). This means that if we estimate the mean brightness of a Cepheid from one random epoch, its uncertainty is $Amp_{W1} / (2\sqrt{2}) \simeq 0.35 \: Amp_{W1}$, since the amplitude of the sinusoidal is half that of the peak-to-peak amplitude of a Cepheid. However, magnitudes from both AllWISE and unWISE Catalogs are based on co-added images taken over multiple epochs. AllWISE was produced from two full-sky scans\footnote{\href{https://wise2.ipac.caltech.edu/docs/release/allwise/}{AllWISE release notes}}, while the observational baseline of unWISE was about five times longer \citep{Schlafly2019}. 
Simulations show, that having two points on a sinusoidal reduces the uncertainty of estimating the mean to $0.5$ of the amplitude, while having ten points to $0.22$ of the amplitude. As a result, the uncertainty in estimating the mean brightness of a Cepheid resulting from pulsations will be $0.25\: Amp_{W1}$ for AllWISE and $0.11 \: Amp_{W1}$ for unWISE.

To be conservative, we combine the photometric errors (Eq.~\ref{eq:sigmaW1}) and the rms scatter of the Cepheid mid-IR light curves in quadrature, as a final measure of uncertainty of the mean brightness measurement:
\begin{equation} \label{eq:sigmaobsW1}
    \sigma_{obs,W1} = 
    \begin{cases}
    \sqrt{ \sigma_{W1}^2 + (0.25 \: Amp_{W1})^2}, & \text{for AllWISE} \\
    \sqrt{ \sigma_{W1}^2 + (0.11 \: Amp_{W1})^2}, & \text{for unWISE} 
    \end{cases}
\end{equation}
\\

\section{Results}
\label{sec:results}

\subsection{Distances}
\label{sec:distances}

The distance to a star (in parsecs) and its uncertainty can be calculated as:
\begin{equation} \label{eq:distance}
\begin{aligned} 
&d = 10^{0.2 \mu + 1}\\
&\sigma_{d} = 0.461 \: d \: \sigma_{\mu} 
\end{aligned}
\end{equation}
where $\mu$ is the extinction-corrected distance modulus:
\begin{equation} \label{eq:mu}
\mu = m - M - A.
\end{equation}
Here $m$ is the observed apparent magnitude, $M$ is the absolute magnitude and $A$ is the extinction.

The observed magnitude $m_{W1}$, or simply $W1$, has been corrected for the following offsets derived for the AllWISE and unWISE catalogs in the previous section: 
\begin{flalign}
\label{eq:offsets}
{\rm AllWISE:} & \: W1 = W1_A + 0.142 & \; {\rm for} \: W1_A < 8.25 \: {\rm mag} \nonumber \\
               & \: W1 = W1_A         & \; {\rm for} \: W1_A \ge 8.25 \: {\rm mag} \nonumber \\
{\rm unWISE:}  & \: W1 = W1_U - 0.024 & 
\end{flalign}

However, we do not use unWISE magnitudes at $W1_U < 4$, nor any unWISE magnitudes with {\em fracflux} values less than 0.2. This removes the unWISE $W1$ photometry from 31 sources. 

The absolute magnitude $M_{W1}$ is calculated from the mid-IR PL relations of \cite{Wang2018} which had been derived specifically for the WISE bands, using 288 Galactic classical Cepheids. We chose the relation determined for 255 fundamental mode pulsators:
\begin{equation}
    M_{W1} = -3.248 \: {\rm log} P - 2.533
    \label{eq:pl}
\end{equation}
with the statistical error $\sigma=0.082$~mag.
The relation does not include the influence of the iron abundance.

\subsubsection{The metallicity dependence of the Cepheid PL relation}

The dependence of the Cepheid PL relations on metallicity has been widely investigated in the past decades, due to its influence on the Hubble constant measurement. Despite the large number of observational studies, the metallicity coefficient $\gamma$ in the PL relations is not tightly constrained and has been found to be positive \citep{Bono2008,Groenewegen2013}, consistent with zero \citep{Udalski2001,Freedman2011,Wielgorski2017,Groenewegen2018,Madore2024}, or negative \citep{Storm2012,Gieren2018,Ripepi2021,Breuval2022,Trentin2024,Bhardwaj2024}, with the latter being more favored in recent years. Moreover, most of the above studies investigate the influence of iron abundance in the optical and near-IR wavelengths, and hardly a few studies extend their coverage to the mid-IR \citep{Freedman2011,Breuval2022}. At the same time, theoretical models indicate that the metallicity coefficient in the PL relations depends on the photometric passband of the observations, such that the metallicity effect becomes smaller at longer wavelengths \citep{Fiorentino2007,Bono2008,Marconi2010}. Indeed, the majority of available Galactic PL relations in the mid-IR show a very small scatter despite not including the metallicity term \citep{Marengo2010,Monson2012,Majaess2013,Wang2018}. However, the only recent study that did include the metallicity term in the mid-IR PL relations \citep{Breuval2022} found it to be negative and similar as in the optical and near-IR bands, although the Cepheid sample with reliable mid-IR data was smaller than in the other bands. Moreover, the mid-IR PL relations from \cite{Breuval2022} were derived for the \textit{Spitzer} $I1$ and $I2$ bands which, while similar to those of WISE $W1$ and $W2$, have slightly different passband characteristics.

Given the uncertain nature of the metallicity coefficient at mid-IR wavelengths and the lack of such relations for the WISE bands, and avoiding the additional uncertainty that would be introduced from using a metallicity from an assumed metallicity gradient, we decided to use the most accurate PL relations from \cite{Wang2018}, which have no metallicity term.  

\subsubsection{Fundamentalizing periods of overtone Cepheids}

\begin{figure}
    \centering
    \includegraphics[width=0.45\textwidth]{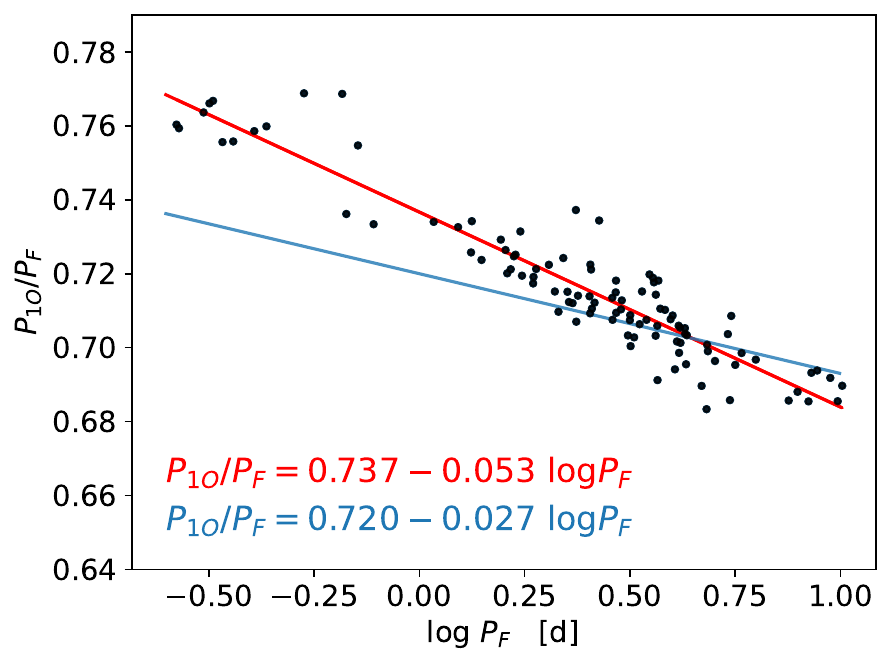}
    \caption{Petersen diagram for 103 Galactic classical Cepheids pulsating simultaneously in the fundamental mode and the first overtone (black dots). The red line is the fit to the data obtained in this work, while the blue line shows the widely used relation from \cite{Alcock1995}, modified by \cite{Feast1997}.}
    \label{fig:fundamentalize}
\end{figure}

Among 3659 Cepheids that are on the P21 list, 2251 pulsate in the fundamental mode (F), 1091 in the first overtone (1O), while the remaining 317 are multi-mode pulsators.  For the purpose of distance calculation, in the case of multi-mode Cepheids, we use the period of the dominant pulsation mode.  Since we want to use the PL relation derived for fundamental mode Cepheids, we need to ''fundamentalize'' periods of Cepheids for which the primary pulsation mode is the first overtone. The transformation between the fundamental mode period, $P_{\rm F}$, and the first overtone period, $P_{\rm 1O}$, is done with the empirical formula
derived from double-mode Cepheids pulsating simultaneously in the fundamental mode and the first overtone (F1O). The traditionally used formula was determined by \cite{Alcock1995} for a small sample of 12 F1O Galactic Cepheids in a narrow period range ($0.3 < {\rm log} P_{\rm F} < 0.8$), and later rewritten by \cite{Feast1997}. The current list of known Galactic Cepheids contains 103 F1O stars in a much larger period range ($-0.6 < {\rm log} P_{\rm F} < 1.0$), we therefore decided to update the relation. The least squares fit to the data gives
\begin{equation}
P_{\rm 1O} / P_{\rm F} = -0.053 \; {\rm log} P_{\rm F} + 0.737
\end{equation}
which can also be written as
\begin{equation}
P_{\rm 1O} / P_{\rm F} = -0.054 \; {\rm log} P_{\rm 1O} + 0.729
\end{equation}
The improvement of the new relation with respect to the one from \cite{Alcock1995} is illustrated in Figure~\ref{fig:fundamentalize}.

\subsubsection{Extinction correction}
\label{sec:ext}

To correct for extinction, S19 used the 3D extinction map $mwdust$ \citep{Bovy2016}, which is a combination of the extinction maps of \cite{Marshall2006} (based on 2MASS data), \cite{Green2015} (based on Pan-STARRS and 2MASS data), and \cite{Drimmel2003} (analytic model for the dust distribution fit to the COBE DIRBE data), depending on the Galactic coordinates and distance.
We follow the same approach although we use an improved version of $mwdust$ which utilizes the map of \cite{Green2019} that takes advantage of the \gaia\ parallaxes for distance calculations and introduces several advancements that refine the maps resolution, with respect to \cite{Green2015}.
The reliability of the $mwdust$ extinction maps has been verified by \cite{Skowron2019a} by finding that $A_{K_S}$ extinction values from $mwdust$ are consistent with those derived from multiband photometry for individual Cepheids.

\begin{figure}
    \centering   \includegraphics[width=0.45\textwidth]{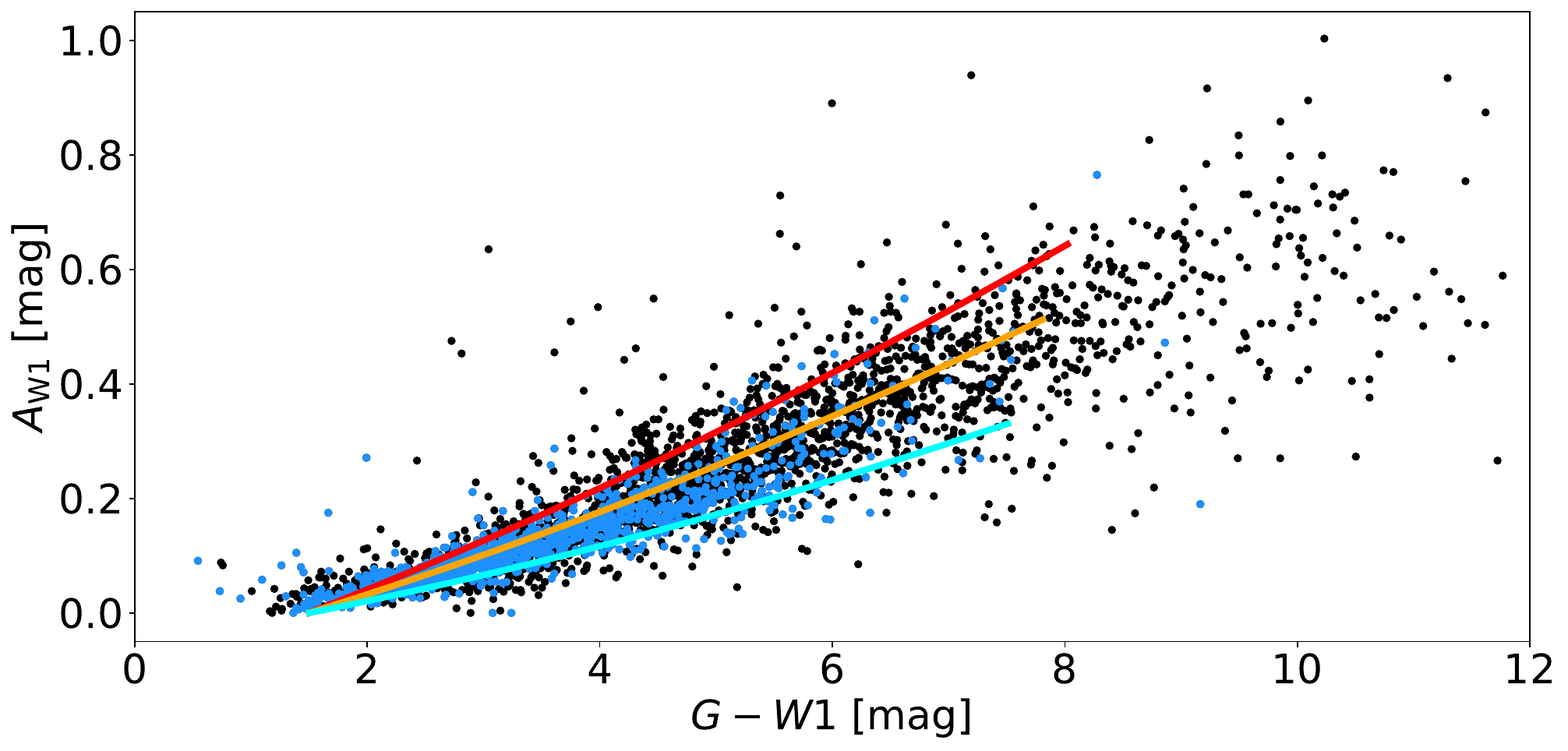}
    \caption{Extinction in $W1$ as a function of $G-W1$ color, shown for the Cepheids from this study as black ($-90^\circ<l<90^\circ$) and blue ($90^\circ<l<-90^\circ$) points. 
    Orange line shows the prediction for $R_{V}=3.1$ assuming the extinction curve from \cite{Gordon2023}, where we adopt a Cepheid intrinsic mean color of $(G-W1)_{0}=1.5$ mag. Cyan and red lines show the predictions for $R_{V}=2.6$ and $R_{V}=3.6$, respectively.
    }
    \label{fig:ddist_DR3-W1_color_theory}
\end{figure}

The 3D extinction map provides extinction values in the $K_S$ band for a given on-sky location and distance. In the process of distance calculation, we extracted the $A_{K_S}$ value for each Cepheid and then transformed it to the WISE $W1$ band using $A_{W1} / A_{K_S} = 0.47$, derived from applying the mid-IR extinction curve from \cite{Gordon2023} \citep[based on data from][]{Gordon2009, Gordon2021, Fitzpatrick2019, Decleir2022} to a source with an effective temperature of 5500$K$ and assuming that the ratio of total-to-selective absorption $R_V = 3.1$. In Figure~\ref{fig:ddist_DR3-W1_color_theory} we show our derived $A_{W1}$ extinctions with respect to the observed $(G-W1)$ colors, and overplot the expectation from our assumed extinction curve, shown with the orange line for a source with an intrinsic color of $(G-W1)_0 = 1.5$ magnitudes. 

We note that $A_{W1} / A_{K_S}$ does not significantly vary over the effective temperature range of Classical Cepheids, but varies between 0.41 to 0.50 for $R_V$ from 2.6 to 3.6, that is at most 20\% over this entire range, or 10\% from the mean value at $R_V=3.1$. 
Therefore, the spread in $A_{W1}$ for a given $G-W1$ color in Figure~\ref{fig:ddist_DR3-W1_color_theory} is primarily due to the variation of $R_V$, as shown with cyan and red lines representing $R_{V}=2.6$ and $R_{V}=3.6$, respectively. Additionally, part of the spread in Figure~\ref{fig:ddist_DR3-W1_color_theory} can be attributed to the width of the instability strip.
Notably, even when considering a much broader range of $R_V$ from 2.6 to 5.5, we arrive at $A_{W1} / A_{K_S}$ between 0.41 to 0.53, corresponding to a variation of about 13\% from its mean value at $R_V=3.1$.

The mean extinction $A_{W1}$ for our sample is 0.22 mag, while 90\% stars have $A_{W1} \leq 0.45$ mag, so that we expect an unaccounted variation of $R_V$ to introduce an uncertainty of $< 0.045$ mag (typically $0.02-0.03$~mag) in our estimated extinctions, which would translate to $<2.1\%$ relative distance uncertainty (typically $1\%$).

In addition to the uncertainty in the estimated extinction due to an unknown variation in $R_V$, there is also the uncertainty coming from the extinction maps.  Unfortunately $mwdust$ does not provide an extinction error for specific coordinates. We find that the majority of the extinction from $mwdust$ for our Cepheids come from either \citet{Marshall2006} or \citet{Green2019}.
The typical $A_{W1}$ uncertainties that come from the Marshall map are about 10\%, with the larger uncertainties being from the brighter nearby sources with a small amount of extinction. However, the typical relative uncertainties cited in the Green map are 3.5\%, which seem overly optimistic, therefore we assume a relative $A_{W1}$ uncertainty of 10\% for both maps. The extinction error from a possible variation of $R_V$ between 2.6 and 3.6 of 10\%, combined in quadrature with this 10\% uncertainty, gives a total possible mean $A_{W1}$ uncertainty of 14\%, which results in 1.4\% maximum relative distance uncertainty for $A_{W1} = 0.22$~mag and 2.9\% for $A_{W1} = 0.45$~mag.

\subsubsection{Final distances and their uncertainties}

The distance and its uncertainty were calculated according to Eq.~\ref{eq:distance}, independently from both AllWISE and unWISE $W1$ photometry. 
Since the extinction term in the distance modulus is distance-dependent, we calculated the distances iteratively, initially assuming no extinction. Then, for the calculated distance, we extracted the $A_{K_S}$ value from $mwdust$, transformed it to $A_{W1}$, and recalculated the distance with the extinction correction. The process was repeated for about $3-4$ iterations until the distance converged on the final value.  

The uncertainty of the distance modulus (Eq.~\ref{eq:mu}) was calculated by summing in quadrature the uncertainties of the observed magnitude $\sigma_{obs,W1}$ (see Eq.~\ref{eq:sigmaobsW1}) and the statistical errors in the mid-IR PL relation (Eq.~\ref{eq:pl}), that include the intrinsic (i.e. cosmic) scatter of the absolute magnitude at a given period, which is 0.082 (see Table~4 in \citealt{Wang2018}):
\begin{equation}
    \sigma_{\mu} = \sqrt{\sigma_{obs,W1}^2 + 0.082^2}
\end{equation}
The uncertainty of our extinction estimate was not included in this error budget and, as we will show in the following section, our uncertainties are likely underestimated. 

\begin{figure}
    \centering
    \includegraphics[width=0.45\textwidth]{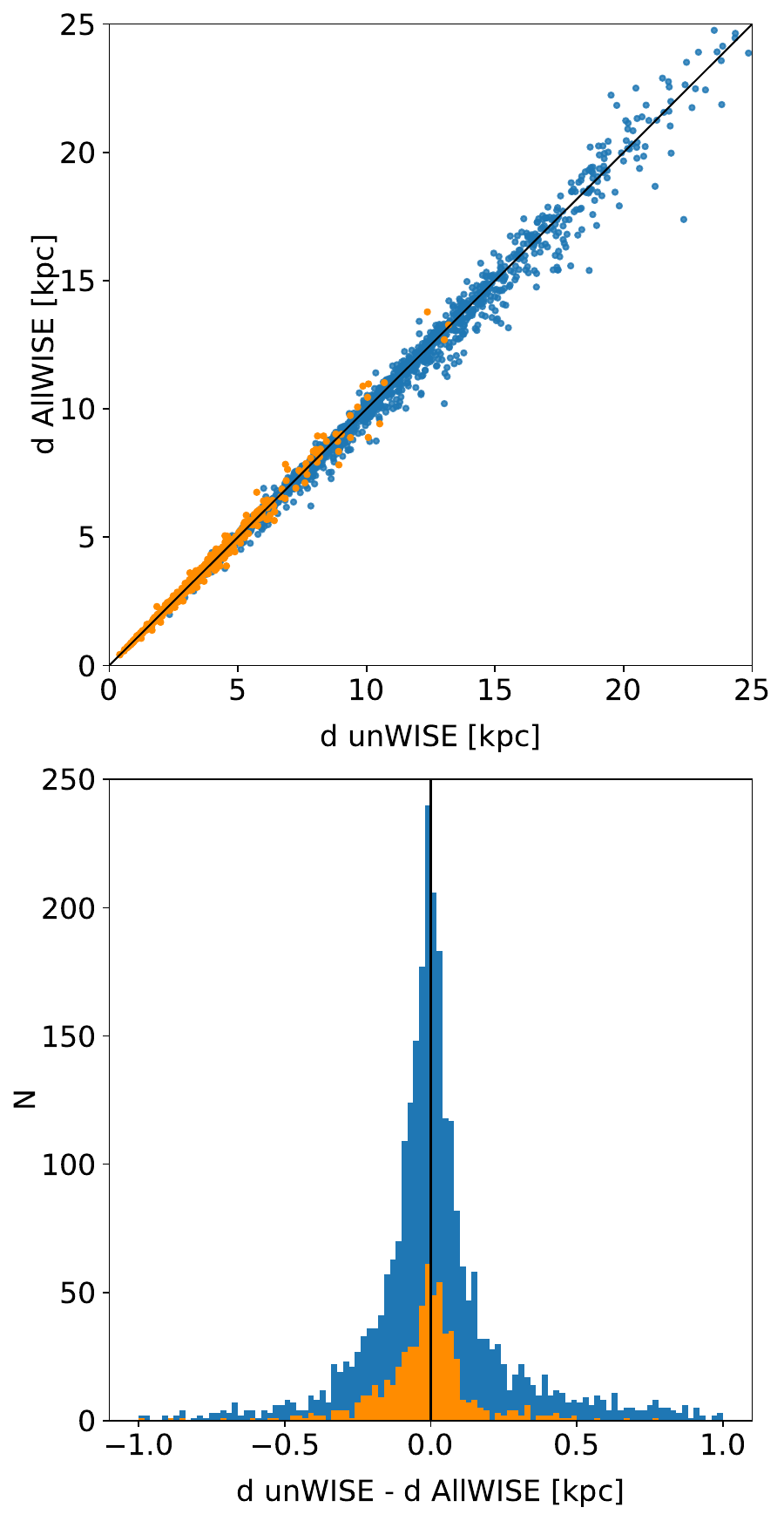}
    \caption{Photometric distances derived from the $W1$ magnitudes for 2780 Cepheids for which both AllWISE and unWISE photometry was available. Orange represents bright Cepheids with $W1_U < 8.25$ mag, while the remaining ones are shown in blue.}
    \label{fig:dist_unWISE_AllWISE}
\end{figure}

As shown in Figure \ref{fig:dist_unWISE_AllWISE},  after the magnitude offsets from Eq.~\ref{eq:offsets} have been applied, both AllWISE and unWISE distances seem to be consistent with each other, even at bright magnitudes (see also Figure \ref{fig:ddist_plx_phot}).
The median differences between AllWISE and unWISE distances are very small, i.e. $-0.002$~kpc for the whole sample, $-0.01$~kpc for Cepheids brighter than $W1_U=8.25$~mag and $0.001$~kpc for fainter than $W1_U=8.25$~mag. The respective standard deviations are $0.472$, $0.205$, and $0.521$~kpc.
We have therefore decided to also derive $W1$-based distance moduli using both the AllWISE and unWISE magnitudes, by averaging the magnitudes when both are available.
The uncertainty of the averaged WISE magnitudes was calculated as half the sum in quadrature of $\sigma_{W1}$ from AllWISE and unWISE.
The final photometric uncertainty was calculated with Eq.~\ref{eq:sigmaobsW1}, using the same amplitude scaling factor as for unWISE data.
We then calculated the distance based on the averaged magnitudes iteratively, as described above.

The final count of Cepheids with $W1$-based distances is 3424, of which 2780 have both AllWISE and unWISE $W1$ magnitudes, while 277 have only AllWISE and 368 have only unWISE $W1$ photometry. In other words, for 2780 Cepheids three different distance estimates are available, while for 645 stars only one distance value based on either AllWISE or unWISE data.

The full set of distances and the distance moduli are available online,
while in Table~\ref{table:cat} we describe the columns in our catalog.

\begin{table*}
\caption{The catalog description.} 
\label{table:cat}
\centering
\begin{tabular}{c|c|c|c|c}
\hline\hline
No. & Column & Description & Unit & Example  \\
\hline 
1 & {\tt survey\_source}  &  Source catalog as reported by P21 & -  & GCVS   \\ 
2 & {\tt survey\_source\_id} &  Source ID in the {\tt survey\_source} catalog &  - & CG\_\_\_\_Cas   \\
3 &{\tt gaia\_id} & \gaia\ source ID  &   - & 429385923752386944  \\ 
4 & {\tt glon}   & Galactic longitude & degrees  & 116.845566   \\ 
5 &{\tt glat}   & Galactic latitude &  degrees & -1.315165   \\     
6 & {\tt mode}   & Mode of variability, from P21 &  -  & F  \\    
7 & {\tt period}  & Fundamentalized Period of variability, from P21 &  days & 4.36540040    \\     
8 & w1\_all   & $W1$ magnitude from AllWISE, after corrections & magnitude  & 7.961  \\  
9 & mu\_all & distance modulus based on AllWISE  &  magnitude &  12.494 \\
10 & dmu\_all & uncertainty in distance modulus based on AllWISE & magnitude & 0.112 \\
11  & d\_all & distance based on AllWISE & parsec & 3154 \\
12 & dd\_all & uncertainty in distance based on AllWISE & parsec &  162\\
13 & a\_all & extinction in $W1$ from AllWISE & magnitude & 0.079 \\
14 & w1\_un & $W1$ magnitude from unWISE & magnitude & 7.989 \\
15 & mu\_un & distance modulus based on unWISE & magnitude & 12.522 \\
16 & dmu\_un & uncertainty in distance modulus based on unWISE & magnitude & 0.090 \\
17 & d\_un  & distance based on unWISE & parsec & 3194 \\
18 & dd\_un  & uncertainty in distance based on unWISE & parsec &  133\\
19 & a\_un & extinction in $W1$ from unWISE & magnitude &  0.079\\
20 & w1\_av & $W1$ magnitude averaged over AllWISE \& unWISE & magnitude & 7.975 \\
21 & mu\_av & distance modulus based on averaged $W1$ & magnitude & 12.508 \\
22 & dmu\_av & uncertainty in distance modulus based on averaged $W1$ & magnitude & 0.089 \\
23 & d\_av & distance based on averaged $W1$ & parsec & 3174 \\
24 & dd\_av & uncertainty in distance based on averaged $W1$ & parsec & 131 \\
25 &a\_av & extinction in $W1$ based on averaged $W1$ & magnitude &  0.079 \\
26 & age & Age estimate & Myr & 128 \\
27 & Q & quality indicator & - & 1.528
  \\  
\hline 
\end{tabular}
\tablecomments{Entries 1-7 (written with different font) are taken from P21, \gaia\, and AllWISE catalogs. The remaining entries had been calculated in this work. The catalog is available in machine readable format.}
\end{table*}

\subsection{Validation of photometric distances}
\label{sec:validation}

\begin{figure}
    \centering
    \includegraphics[width=0.49\textwidth]{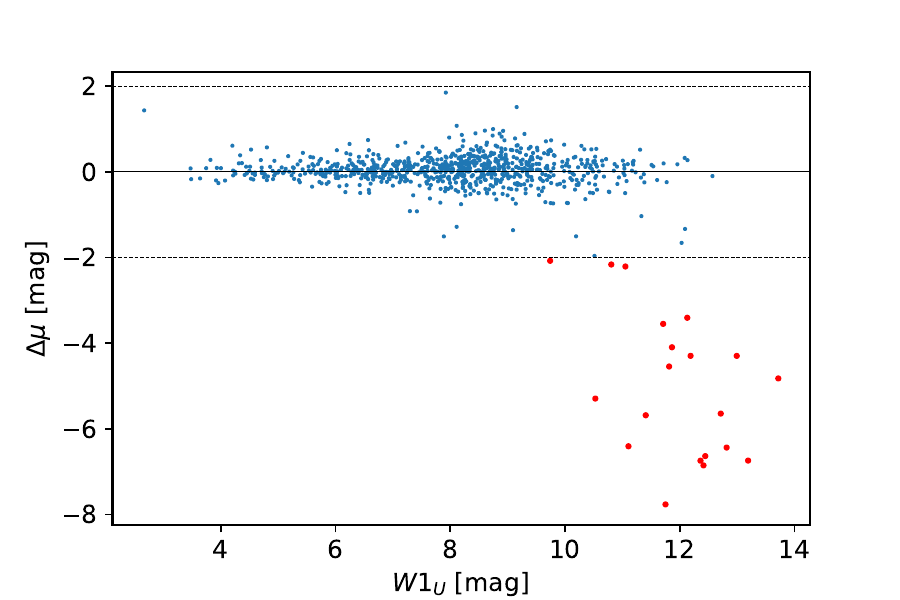}
    \caption{Difference in the $W1$-based photometric and parallax-based astrometric distance moduli ($\Delta \mu$) with respect to the unWISE $W1$ magnitudes for the subset of Cepheids with $\sigma_\varpi/\varpi < 0.1$. Dashed horizontal lines show the $\Delta \mu = \pm 2$ magnitude limits, and the red points mark 20 Cepheids falling outside these limits.}
    \label{fig:ddmod_W1}
\end{figure}

To empirically check the quality of our distances and distance uncertainties we use the subset of 930 Cepheids with \gaia\ parallaxes of good quality ($\sigma_\varpi/\varpi < 0.10$ and $ruwe < 1.4$), of which 910 have $W1$-based distances.
Here we will compare the distance modulus derived from the mid-IR photometry, $\mu_{W1}$ (Eq.~\ref{eq:mu}), to the distance modulus derived from the parallax, that is,
\begin{equation}
    \mu_\varpi = -5 \log(\varpi) + 10
    \label{eq:astrodm}
\end{equation} 
where $\varpi$ is the parallax in mas.
Figure \ref{fig:ddmod_W1} shows the difference in the distance moduli $\Delta \mu = \mu_\varpi - \mu_{W1}$ as a function of the unWISE $W1$ magnitudes.  While the majority of the sources show overall consistent distances, there is a conspicuous set of sources whose $\Delta \mu < 2$ mag.
In particular, we find 20 stars classified as Cepheids according to their light curves, but whose astrometric distances show they cannot be Cepheids. All of these stars are classified as first-overtone oscillators.  We inspected the light curve of these sources and found that, after including more recent photometric observations, a number now appear to be spotted stars, while others cannot be clearly classified. However, at least five still appear to be Cepheids, although this classification may change as more data are available (P. Pietrukowicz, private communication).  

Expanding to 1443 Cepheids with $\sigma_\varpi/\varpi < 0.20$ and $ruwe < 1.4$, we find an additional 18 stars (38 total) with inconsistent astrometric distances, again mostly classified as first overtone oscillators. 
In summary, our sample of Cepheids from P21 may be contaminated by miss-classified objects at the level of 2 to 3 percent.  These are the same sources noted by \citet{Ripepi:2023} (see their section 6.2) in their comparison of the \gdrthree\ classifications with respect to P21 classifications. Indeed, since \gdrthree\ classifications are informed by the parallax, only three of the 20 Cepheids with inconsistent distances are classified as Cepheids in \gdrthree.  Rather than remove these problematic sources from our Cepheid sample, below we provide a quality indicator that measures the consistency of the astrometric and photometric distances. 

To evaluate the photometric distance uncertainties in an empirical fashion, we assume that the parallax uncertainties reported in \gdrthree\ are correct, and we generate a set of hypothetical "true" parallaxes by bootstrapping (with random draw and replace) the $\{ \varpi, \sigma_\varpi \}$ pairs from our sample of 910 sources with good parallaxes and consistent distances, and using the parallax as a true parallax $\varpi_{true}$. We then produce a resampled "observed" parallax 
\begin{equation}
    \varpi = \varpi_{true} + \epsilon_\varpi - \varpi_{0}, 
\label{eq:parallax}
\end{equation}
where $\varpi_{0}$ is the parallax zeropoint (i.e. offset), and $\epsilon_\varpi$ is a random normal deviate with a standard deviation that is the sum in quadrature of the associated $\sigma_\varpi$, and an additional $\sigma_0$ to account for the variation of the parallax offset for our subsample \citep[][hereafter L21]{Lindegren2021b}. 
That is, we assume that the parallax offset has a mean value $\varpi_{0}$, and varies from this as a Gaussian with a standard deviation of $\sigma_0$.  In short, the set of 910 resampled parallaxes $\varpi$ is sampled from the distribution $\mathcal{N}(\varpi\, |\, \varpi_{true} - \varpi_0,\sigma^2_\varpi + \sigma^2_0)$ where $\mathcal{N}(x\,|\,\langle x \rangle,\sigma^2)$ is a normal distribution with mean $\langle x \rangle$ and standard deviation $\sigma$.
Our resulting distribution of resampled parallaxes will of course be smoother than the actual data since we are taking $\varpi_{true}$ from a bootstrap of the observed parallaxes, but we are not trying to model or replicate the distribution of parallaxes.  Our procedure is a simple way to generate a "realistic" distribution of model observations with (conveniently) similar correlations as is in the actual data, without needing to explicitly model these correlations and the parallax error distribution.  For example, more distant (fainter) sources will have larger errors in our bootstrap sample just as in the real data since we drew $\{ \varpi, \sigma_\varpi \}$ pairs from the data itself. 

What we are trying to replicate is the distribution of $\Delta \mu$, on the assumption that the photometric and astrometric errors are independent of one another.  To generate a distribution of $\Delta \mu$ we derive $\mu_\varpi$ from our resampled parallaxes according to equation \ref{eq:astrodm}, using the true parallax we generate an observed photometric distance modulus assuming random Gaussian errors in the photometry, that is, the modeled photometric distance modulus is $\mu = -5 \log(\varpi_{true}) + 10 + \epsilon_{\mu}$ where $\epsilon_{\mu}$ is drawn from a normal distribution $\mathcal{N}(0,\sigma_{\mu})$. Since our subsample of Cepheids with good parallaxes have magnitudes $W1 < 12$, we assume a single value for the uncertainty of our photometric distances, $\sigma_{\mu}$, for our subsample and adjust $\sigma_{\mu}$, together with $\varpi_{0}$ and $\sigma_{0}$, to reproduce our observed distribution of $\Delta \mu$.  

To summarize, our model has three parameters, ($\varpi_0, \sigma_0, \sigma_\mu$).  We generate a model $\Delta \mu$ distribution by using a single set of $\{ \varpi_{true}, \sigma_\varpi \}$ with the same size (910) as our data set, but oversampling the noise by a factor of 100, binning the resulting $\Delta \mu$ distribution as done with the data set (40 bins in the range [-2,2]) and dividing the counts by 100.  We perform an adjustment of the three parameters ($\varpi_0, \sigma_0, \sigma_\mu$) by minimizing the $\chi^2$ of the residuals of the counts in the bins, that is,
\begin{equation}
    \chi^2 = \sum_i \frac{(n_i - m_i)^2}{\sigma_i^2},
\end{equation}
where the sum is over the bins where $n_i \neq 0$, $n_i$ and $m_i$ being the bin counts for the data and the model respectively, and $\sigma_i$ is the uncertainty in the counts for the binned data, assuming Poissonian errors ($\sigma_i = \sqrt{n_i}$).  The minimization is done using the scipy.optimization.minimization package employing the Powell method.  We arrive at a final estimate of the parameters and their uncertainties by repeating this procedure 100 times with 100 different bootstrap samples of $\{ \varpi_{true}, \sigma_\varpi \}$, then taking the mean and standard deviation of the 100 parameter estimates as their final estimate and uncertainty. 
This above procedure results in the values $(\varpi_0, \sigma_0, \sigma_\mu)$ reported in the second column of Table \ref{tbl:offsets}.

\begin{figure}
    \centering
    \includegraphics[width=0.49\textwidth]{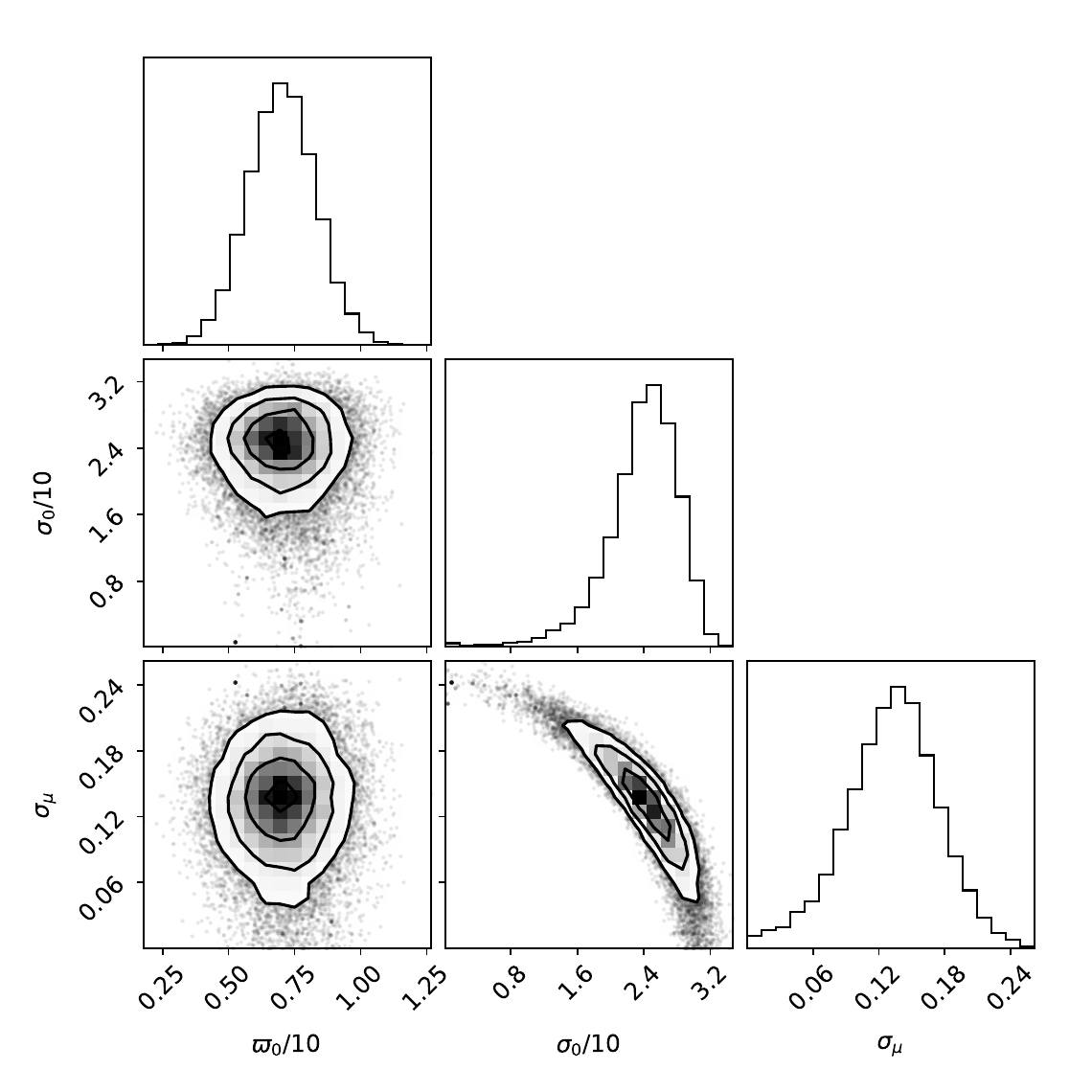}
    \caption{Corner plot of the distribution of possible model parameters $(\varpi_0, \sigma_0, \sigma_\mu)$ based on the MCMC samples after maximizing the likelihood of the $\Delta \mu$ distribution with respect to the modeled distribution.}
    \label{fig:ddmod_corner_maxlike}
\end{figure}

\begin{figure}
    \centering
    \includegraphics[width=0.49\textwidth]{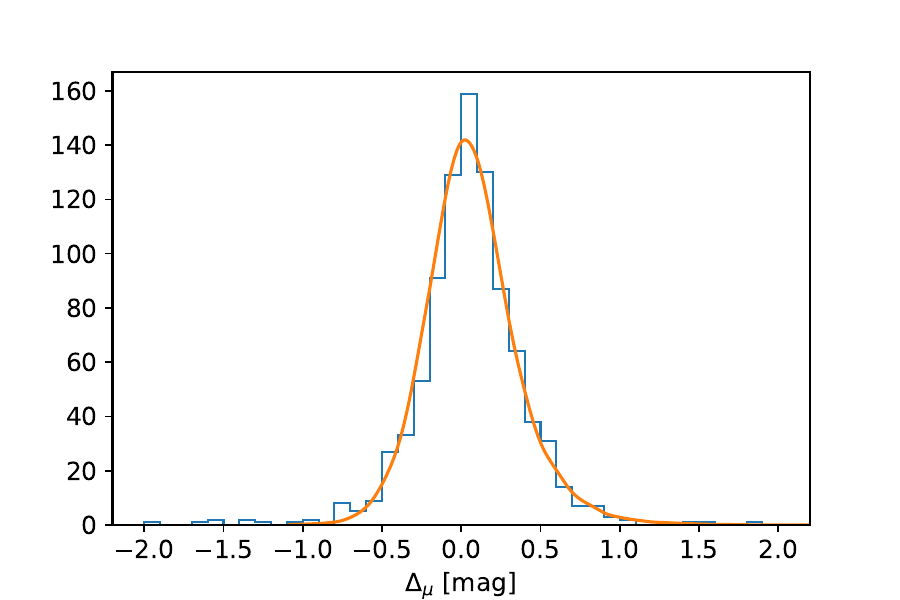}
    \caption{Distribution of the difference in the $W1$-based photometric and astrometric distance moduli ($\Delta \mu$) as seen in the subset of 910 Cepheids with good parallaxes (histogram), and the probability density function (renormalized to the binned counts) of such a sample modeled with the parameters estimated via maximum likelihood  (orange line). See text for details.}
    \label{fig:ddmod_hist}
\end{figure}

To find a more refined estimate of our parameters that does not rely on binning the data, we instead maximize the Likelihood of our dataset 
$\{ \Delta \mu_i \}$ using Markov chain Monte Carlo (MCMC), as implemented in the \textsc{emcee} package \citep{Foreman-Mackey:2013}.  We again generate a model distribution of $\Delta \mu$ from a single set of $\{ \varpi_{true}, \sigma_\varpi \}$ bootstrapped from our data set, but oversampling the noise by a factor of 20.  From this distribution we derive a probability distribution function (PDF) using a normalized kernel density estimator from the \textsc{sklearn} package \cite{scikit-learn}, using a Gaussian kernel with a bandwidth of 0.05 mag and sampled with 1000 points in $\Delta \mu$ over the full range of our stochastic sample generated from our bootstrap sample, typically covering the range [-1,2]. The probability of each data point $\Delta \mu_i$ is found from the PDF by linear interpolation. 
We then find the log likelihood ($\log L$) of our dataset $\{ \Delta \mu_i \}$ by summing the log probabilities of the individual points.  As our dataset does include outliers that fall outside the range of our stochastically modeled PDF but contribute to $\log L$, we found it necessary to "stabilize" $\log L$ by setting a minimum value for the probability to $10^{-6}$. 
We estimate our parameters by using MCMC with 20000 steps and 24 walkers, and using the $\chi^2$ estimated values above as a first guess. In addition we found it necessary to limit the range of possible parameter values explored by the MCMC by setting the $log L$ to negative infinity if the explored value of any of the three parameters falls outside the prescribed range. These ranges are reported in the last column of Table \ref{tbl:offsets}, which summarizes the resulting best estimates and their uncertainties from the MCMC samples after removing the first 4000 steps to allow for burn in. Figure 
\ref{fig:ddmod_corner_maxlike} shows the resulting corner plot of possible parameter values and Figure \ref{fig:ddmod_hist} shows the observed $\Delta \mu$ distribution and its modeled PDF using these final parameter estimates from the MCMC maximizing the Likelihood.

\begin{table}
\caption{Estimated parameters for parallax zeropoint, parallax zeropoint variance and photometric uncertainties.}
\label{tbl:offsets}
\centering
\begin{tabular}{c c c c}
\hline\hline
parameter & $\chi^2$ & Likelihood & Range \\
\hline
  $\varpi_0$ & $7.9 \pm\ 0.6 \: \muas$ &  $7.0 \pm 1.3 \: \muas$  & $[-50,50]\: \muas$\\
  $\sigma_0$ & $23.7 \pm\ 3.8 \: \muas$ & $23.8 \pm 4.3 \: \muas$  & $[0,50] \: \muas$\\  
  $\sigma_\mu$ & $0.13 \pm\ 0.03$~mag & $0.13 \pm 0.04$~mag & $[0,0.5]$~mag\\
\hline
\end{tabular}
\end{table}
  
We note that the offset of the mode of the $\Delta \mu$ distribution from zero, as well as the asymmetry of the distribution, can only be attributed to the astrometric distance moduli errors, the offset being primarily due to the parallax offset $\varpi_0$ and the asymmetry due the parallax errors.  Our estimated parallax offset is comparable with other estimates in the literature, as discussed below in Section \ref{discussion}. To summarize, we estimate the uncertainty of our photometric distance moduli $\sigma_\mu$ to be $0.13 \pm 0.04$ mag, that is, to have relative distance errors of less than 6\% in the magnitude range considered. Meanwhile the mean uncertainty of the distance moduli of this subsample of stars, derived in the previous section, is 0.09  mag.  
Our verification using the measured parallaxes thus suggest that our estimated errors on the distance moduli are somewhat under-estimated, at least for this subset of the nearest Cepheids. 
This under-estimation of the uncertainties cannot be attributed solely to the neglected extinction uncertainties, as the mean $A_{W1}$ of this subsample is only 0.12 mag so that even a 50\% uncertainty in the extinction would only contribute an additional 0.02 mag to the distance modulus uncertainties. However, it is worth noting from Figure \ref{fig:ddmod_corner_maxlike} that $\sigma_0$ and $\sigma_\mu$ are correlated and, more fundamentally, our analysis is based on the assumption that there is no systematic error in the magnitudes. 

\begin{figure}
    \centering
    \includegraphics[width=0.48\textwidth]{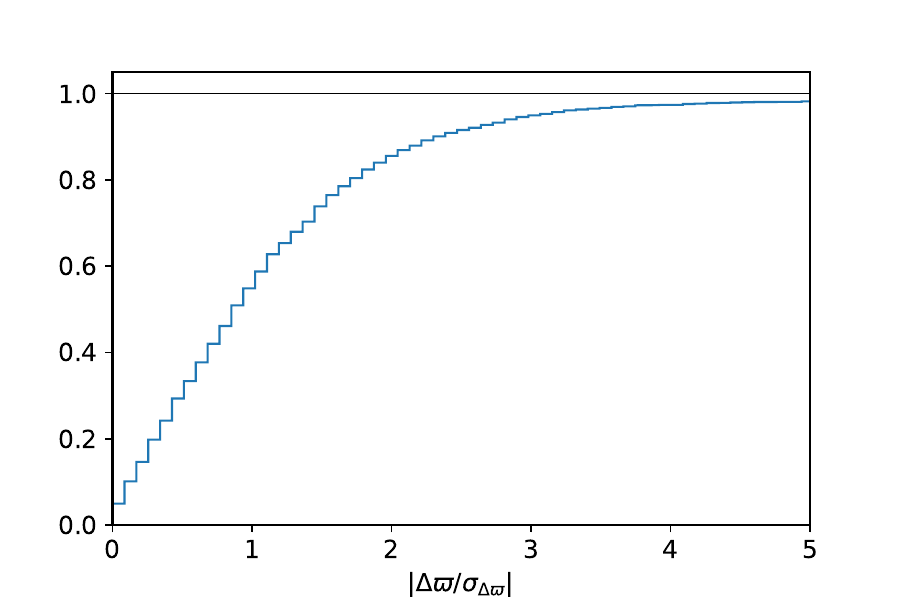}
    \caption{Cumulative distribution of the normalized difference in the $W1$-based photometric and parallax-based astrometric parallaxes ($Q = |\Delta \varpi / \sigma_{\Delta \varpi}|$).}
    \label{fig:dplx_norm_cdist}
\end{figure}

Now that we have characterized our photometric distance modulus errors, we can provide a quantification of the consistency of the astrometric and photometric distances, or rather, the consistency of the astrometric and photometric parallaxes, where the photometric parallax (in mas) is 
\begin{equation}
    \varpi_\mu = 10^{-(\mu - 10)/5}\,.
\end{equation}
We can then consider the normalized difference in the parallaxes as a quality indicator, that is
\begin{equation}
    Q = \left| \frac{\Delta \varpi}{\sigma_{\Delta \varpi}} \right| \,, 
\end{equation}
where $\Delta \varpi = \varpi - \varpi_\mu$, and $\sigma_{\Delta \varpi}$ is the sum in quadrature of $\sigma_\varpi$, $\sigma_0$, and the uncertainty of the photometric parallax, $\sigma_{\varpi_{\mu}}$.  We estimate the later as
\begin{equation}
   \sigma_{\varpi_{\mu}} = \frac{\varpi_+ - \varpi_-}{2}
\end{equation}
where
\begin{equation}
    \varpi_\pm = 10^{-(\mu \mp \sigma_\mu - 10)/5}\,.
\end{equation}
We find that $Q < 5$ removes the 20 anomalous sources with good parallaxes and inconsistent astrometric/photometric distance moduli ($\Delta \mu < -2$ mag) mentioned above.  
Applying this criteria to the complete sample of Cepheids removes an additional 43 sources, that is 63 out of 3424 (1.8\%) of our sample with distance moduli. In Figure \ref{fig:dplx_norm_cdist} we show the cumulative distribution of $Q$, which can be used as a quality indicator for selecting subsamples.

To test whether our uncertainties are underestimated due to neglecting extinction uncertainties, we also checked whether $Q$ was correlated with extinction $A_{W1}$.  As the uncertainty in extinction should be proportional to extinction, higher $Q$ values should result from systematically underestimating the uncertainties with higher extinction.  However we found no evidence of such a correlation between $Q$ and $A_{W1}$.

\subsection{Comparison to mid-IR based distances}

In this paper we calculated distances to classical Cepheids using the same approach as in S19, therefore we should expect our results to be comparable. S19 provide distances to 2390 Galactic Cepheids, of which 2219 stars have new distances determined in this work, meaning that 171 stars from S19 are not in our sample. There are two reasons behind this discrepancy. One is that the classical Cepheid list used by S19 contained stars which have since been reclassified as other types of variable stars and removed from the P21 list. The second reason is that S19 use both the \textit{Spitzer} and WISE data and average individual distances from four WISE and four \textit{Spitzer} bands (or as many as are available), while we use only the WISE $W1$ photometry, therefore some of these stars are the ones that only have mid-IR \textit{Spitzer} measurements, and no WISE photometry.

\begin{figure}[t]
    \centering
    \includegraphics[width=0.99\columnwidth]{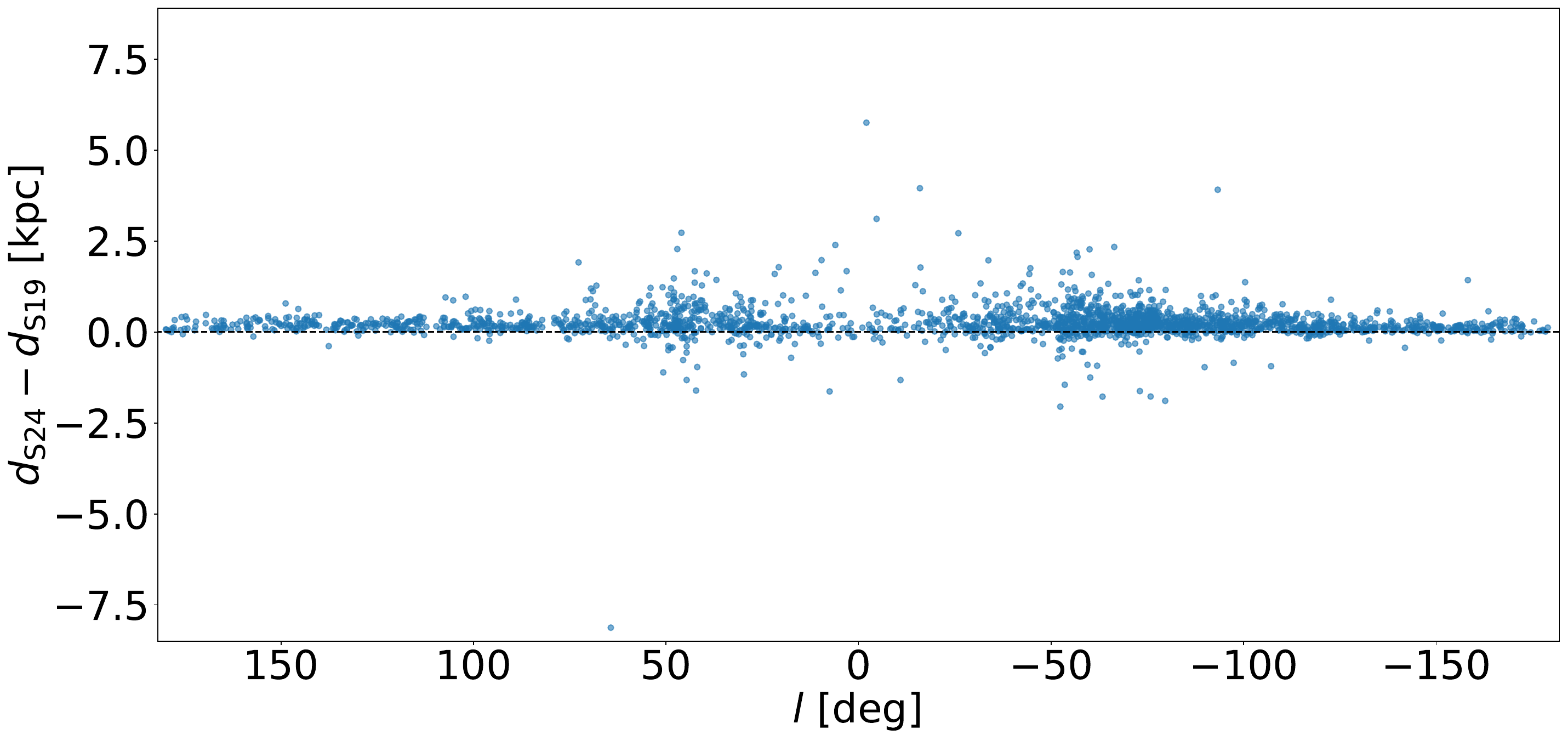}
    \caption{The difference in Cepheid distances from this study (S24) and S19, as a function of Galactic longitude, for 2096 classical Cepheids. The mean offset is $242 \pm 446$~pc.} 
    \label{fig:distances_skowron19_skowron_longitude}
\end{figure}

The distance difference against Galactic longitude is presented in Figure~\ref{fig:distances_skowron19_skowron_longitude}. The overall agreement is good, as expected, especially when compared with Figure~\ref{fig:distance_comparison_intro}. There is however a mean offset of $242 \pm 446$~pc which could be due to several differences between these studies. One, already mentioned above, is the use of observations in all available mid-IR bands by S19. In addition, they utilize the WISE photometry from the AllWISE Multiepoch Photometry Table, which provides time-resolved observations.
The second difference is that we correct for the offsets between the AllWISE and unWISE catalogs (see Section~\ref{sec:photometry} and Eq.~\ref{eq:offsets}). This results in a large mean offset in $W1$ magnitude between this study and S19: $W1_{\rm S24} - W1_{\rm S19} = 0.155 \pm 0.109$~mag for $W1<8.25$~mag, while only $0.005 \pm 0.045$~mag for $W1>8.25$~mag. However, when we exclude stars brighter than $W1=8.25$~mag from the distance comparison, the mean distance offset is still $238 \pm 506$~pc, which is similar to the one for the full sample. 

\begin{figure}
    \centering
    \includegraphics[width=0.45\textwidth]{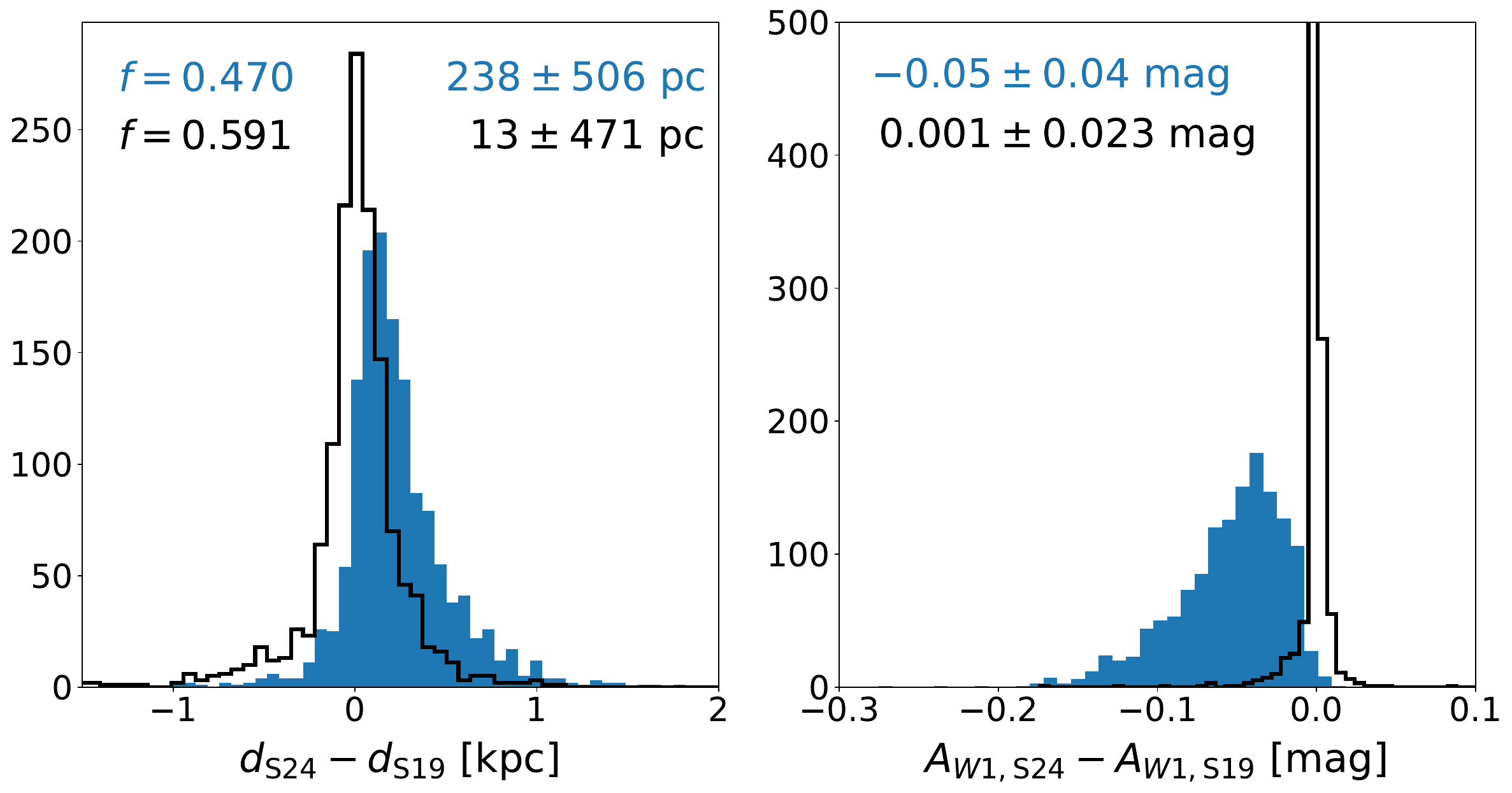}
    \caption{The distribution of differences in distance (left) and extinction (right) between this study and S19, for Cepheids with $W1>8.25$~mag. In blue we plot the distance/extinction difference when using the 0.470 conversion factor and \cite{Green2019} maps, while the black line, with the use of 0.591 factor and \cite{Green2015} maps. Mean offsets are marked on the plots with corresponding colors.} 
    \label{fig:4histograms}
\end{figure}

Another major difference between S19 and this study is that we use the newer version of 3D extinction map $mwdust$ which utilizes the map of \cite{Green2019} over the map of \cite{Green2015}, and a $0.470$ conversion factor between $A_{K_S}$ and $A_{W1}$ instead of 0.591 used by S19. For the purpose of the comparison we repeated our distance estimation procedure using $A_{W1}/A_{K_S}=0.591$ and \cite{Green2015} extinction maps for 1408 Cepheids with $W1>8.25$~mag. The difference in both distance and $A_{W1}$ extinction is plotted in the left and right panels of Figure~\ref{fig:4histograms}, respectively. In blue, we show the distance/extinction difference when using the 0.470 conversion factor and \cite{Green2019} maps, while in black, with the use of 0.591 factor and \cite{Green2015} maps. As anticipated, our results are more consistent with those of S19 when using $A_{W1}/A_{K_S}=0.591$ and \cite{Green2015} extinction maps, the mean difference in distances being only $13$ instead of $238$~pc. We can therefore conclude that new distances differ from S19 primarily due to the change in the estimation of the extinction. 

\subsection{Comparison to distances based on the \gaia\ Wesenheit index}

\cite{GaiaCollaboration2023} estimated distances to 3306 Galactic classical Cepheids using photometric data from \gdrthree\ and period--Wesenheit--metallicity relations calibrated specifically for the {\em Gaia} bands \citep{Ripepi2022}, following the assumption that the use of the reddening-free period--Wesenheit relation does not require an extinction correction. These include 269 stars from the literature which were not classified as classical Cepheids by \gdrthree, as well as 269 stars which have since been reclassified by P21 as other types of variable stars.

\begin{figure}[t]
    \centering
    \includegraphics[width=0.99\columnwidth]{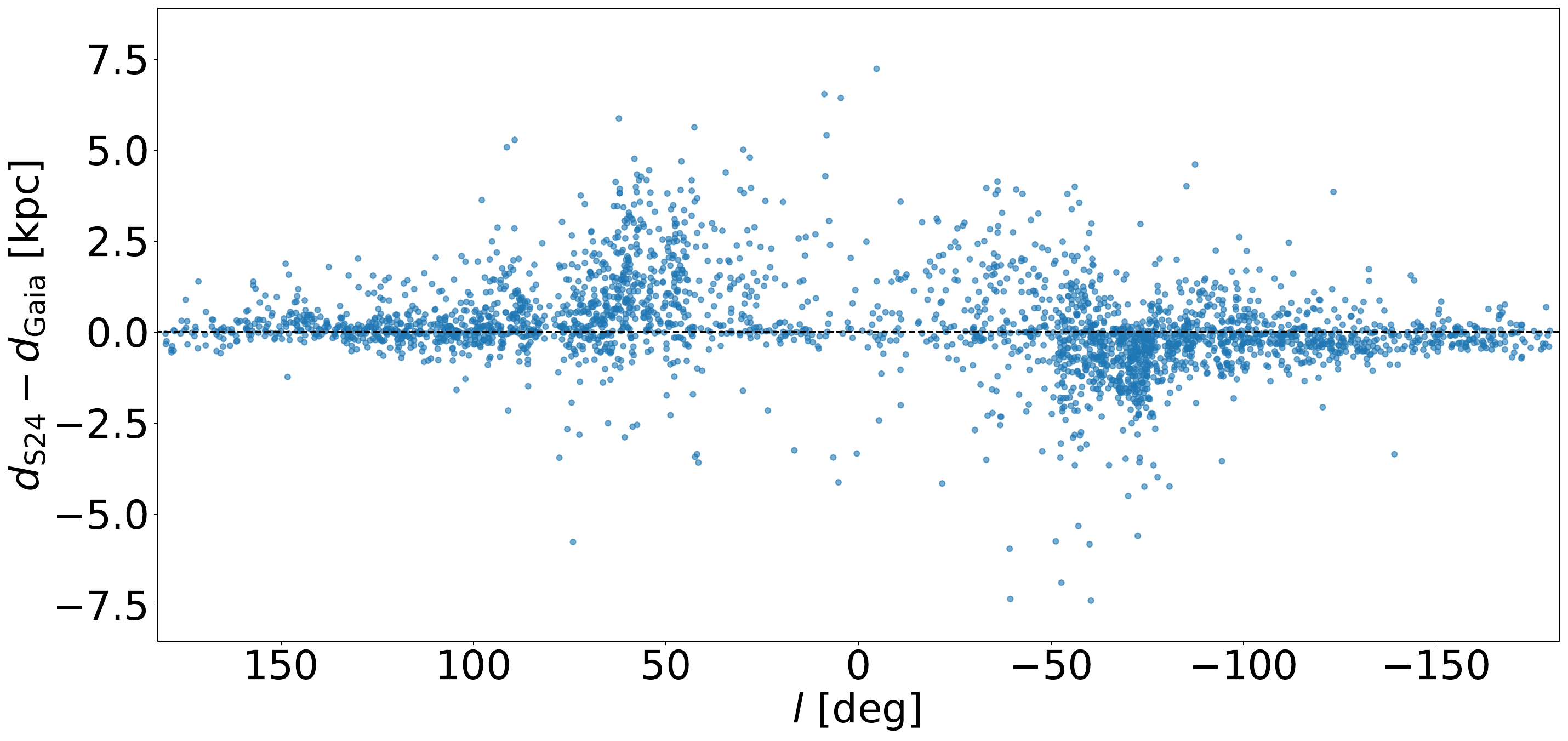}
    \caption{The difference in Cepheid distances from this study (S24) and based on \gaia\ (optical) Wesenheit magnitudes as a function of Galactic longitude, for 2910 classical Cepheids. The mean offset is $179 \pm 1293$~pc.}
    \label{fig:ddist_DR3-W1_glong}
\end{figure}

\begin{figure}
    \centering
    \includegraphics[width=0.4\textwidth]{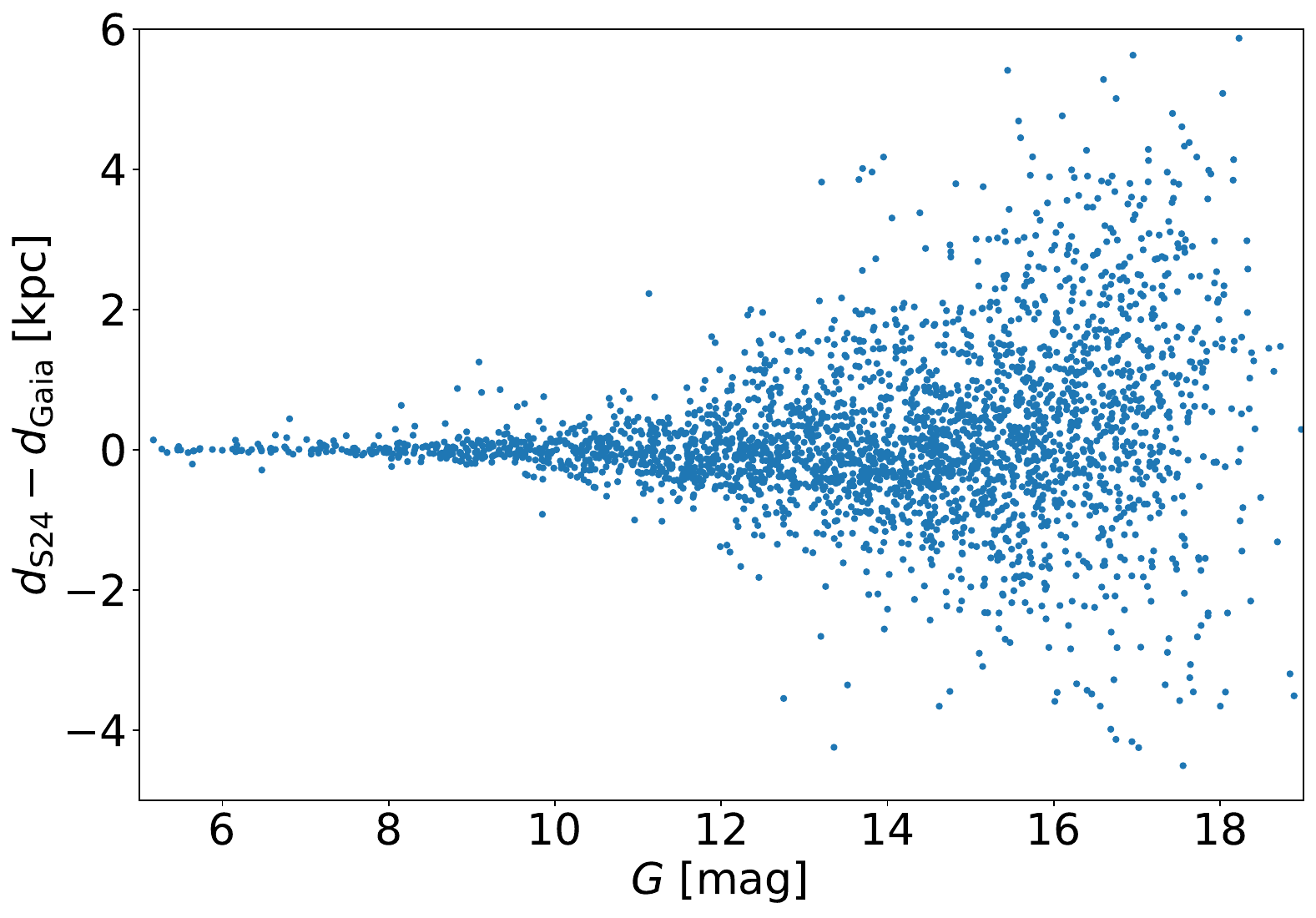}
    \includegraphics[width=0.4\textwidth]{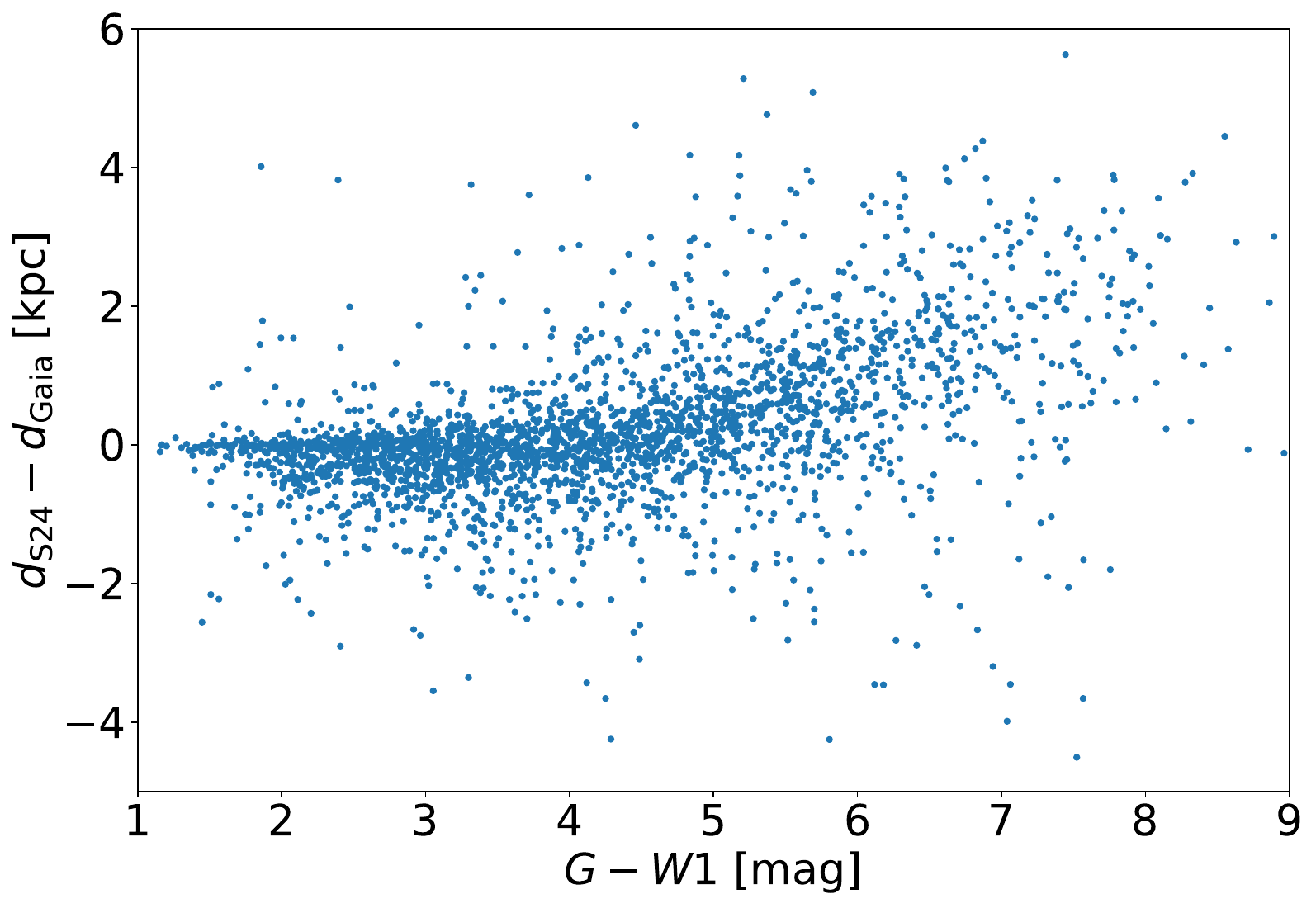}
    \caption{Differences in the distances from this study and based on \gaia\ (optical) Wesenheit magnitudes, as a function of $G$ magnitude (top) and $G-W1$ color (bottom).}
    \label{fig:ddist_DR3-W1_color}
\end{figure}

Figure \ref{fig:ddist_DR3-W1_glong} shows the distance difference between this study and \cite{GaiaCollaboration2023}, plotted against Galactic longitude, for 2910 Cepheids that have distances from both studies.
As expected, the largest discrepancies are observed toward the inner Galaxy, where the extinction is highest -- the mean offset is $179$~pc (with the dispersion of $1293$) for the whole sample of 2910 stars, while it is $25$~pc (with the dispersion of  $544$) for the 885 Cepheids located away from the Galactic Center ($l<-100\deg$ or $l>100\deg$), showing that both methods give similar results in regions of low extinction.
However, the discrepancies between the two distance determinations become larger as stars become redder (see Figure~\ref{fig:ddist_DR3-W1_color}), showing that these discrepancies are due to the difference in how extinction is accounted for in these two methods.

In Section~\ref{sec:ext} we demonstrated that extinction errors from $mwdust$ maps combined with those from the possible variation of $R_V$ between 2.6 and 3.6 could results in not more that 2.9\% distance uncertainties for our sample of Cepheids, with a mean value of 1.4\%.
On the other hand, distance uncertainties due to $R_V$ deviating from the assumed value of 3.1 are much larger when optical Wesenheit magnitudes are used for distance determination.  For example, in \citet{GaiaCollaboration2023}, Wesenheit magnitudes from the \gaia\ photometry were used, according to the prescription
\begin{equation}
W_G = G - R_{Gaia} (G_{BP} - G_{RP}),
\end{equation}
where the color coefficient $R_{Gaia}=1.9$ is a consequence of the assumed extinction law \citep{Cardelli:1989}, adopting a typical value of $R_V = 3.1$. Assuming the extinction curve of \citet{Gordon2023} instead, would only change this coefficient slightly (to about 2.0), as all recent extinction curves are very similar in the optical.  However, as mentioned above, $R_V$ values can vary between 2.6 and 3.6 near the Galactic plane. 
In this range, the color coefficient $R_{Gaia}$ would vary between about 1.7 and 2.2. For a Cepheid with an observed $(G_{BP} - G_{RP})$ color of 2.5 magnitudes (the mean observed color of our Cepheid sample), the Wesenheit magnitude $W_G$ would then vary by $\pm 0.6$ magnitudes, introducing a relative distance uncertainty of 30\%, which is an order of magnitude larger than the one resulting from using mid-IR based distances.

\begin{figure}
    \centering
    \includegraphics[width=0.99\columnwidth]{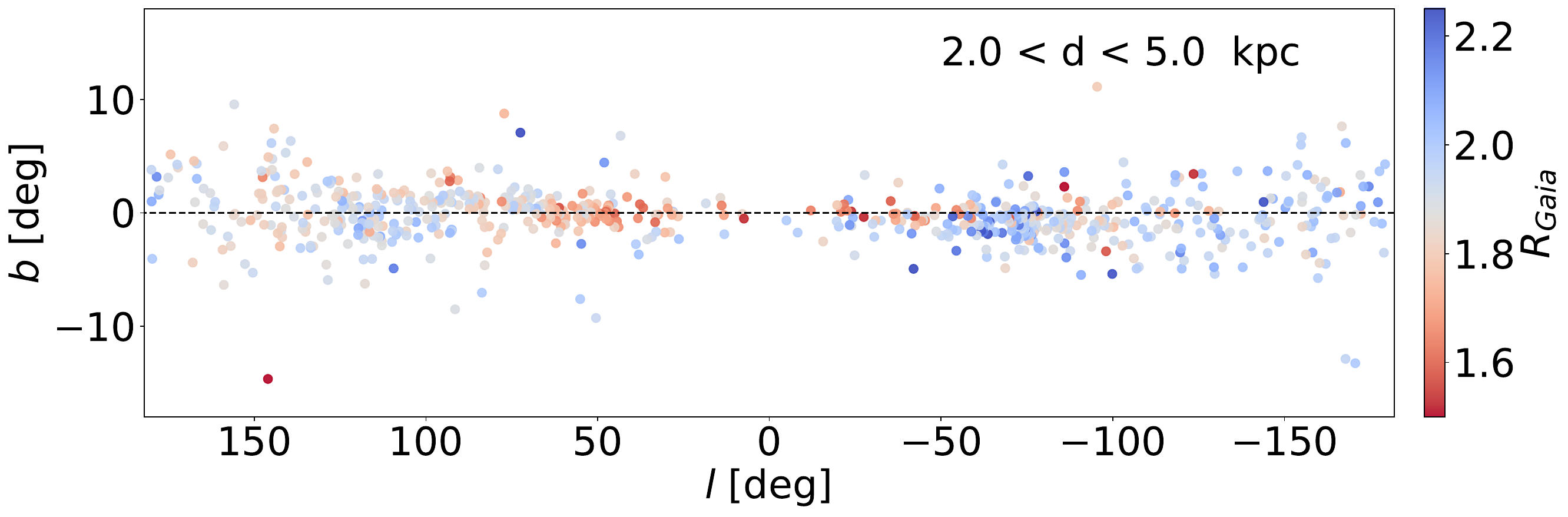}
    \caption{The distribution of 695 Cepheids within the distance range $2-5$ kpc in Galactic coordinates, for which both $d_{S24}$ and $d_{Gaia}$ are available. The color represents the expected value of $R_{Gaia}$ estimated from the distance difference $(d_{S24} - d_{Gaia})$ assuming that $d_{S24}$ distances are correct. To be compared with Fig.~2 from \cite{Zhang2025}.}
    \label{fig:RGaia}
\end{figure}

As an additional test, we reversed the problem and investigated how the distance differences that we observe $(d_{S24} - d_{Gaia})$ could be due to a variation in the $R_{Gaia}$ coefficient. For the purpose of this exercise we assumed that our Cepheid distances $d_{S24}$ are correct, and for each Cepheid we calculated the value of $R_{Gaia}$ that the Cepheid should have in order for its Wesenheit-based distance $d_{Gaia}$ to match the $d_{S24}$ distance.
Figure~\ref{fig:RGaia} shows the distribution of $R_{Gaia}$ calculated for 695 individual Cepheids within the distance bin [2, 5] kpc, so that it can be easily compared to the distribution of $R_V$ from \cite{Zhang2025} in the same distance bin, shown in the very bottom of their Fig.~2. While we do not expect a perfect agreement, it is readily visible that $R_{Gaia}$ follows the same trend as $R_V$. That is, in regions where $R_V$ is low (e.g. $-50\deg < l < 50\deg$), $R_{Gaia}$ is also low, and in regions where $R_V$ is high (e.g. $-100\deg < l < -50\deg$), $R_{Gaia}$ is also high. Also, the range of variation in $R_{Gaia}$ required to account for the observed differences is reasonably modest.
This is yet another argument that the variation in $R_V$ is the main reason behind the distance differences that we observe and further supports moving to the mid-IR regime, where the extinction is not only smaller, but also less sensitive to changes in $R_V$.

\subsection{Cepheid ages}

For age estimates of our Cepheids we use the period-age relation derived by \citet{Anderson2016}, who take into account the effect of rotation. Following the authors recommendations, we assume the initial rotation rate $\omega = 0.5$ which denotes the ratio of the initial to critical rotation velocity, as well as an unknown location within the instability strip and an unknown crossing number, that is, whether this is the first, second or third crossing of the instability strip (Table~4 in \citealt{Anderson2016}).

\begin{figure}
    \centering
    \includegraphics[width=0.45\textwidth]{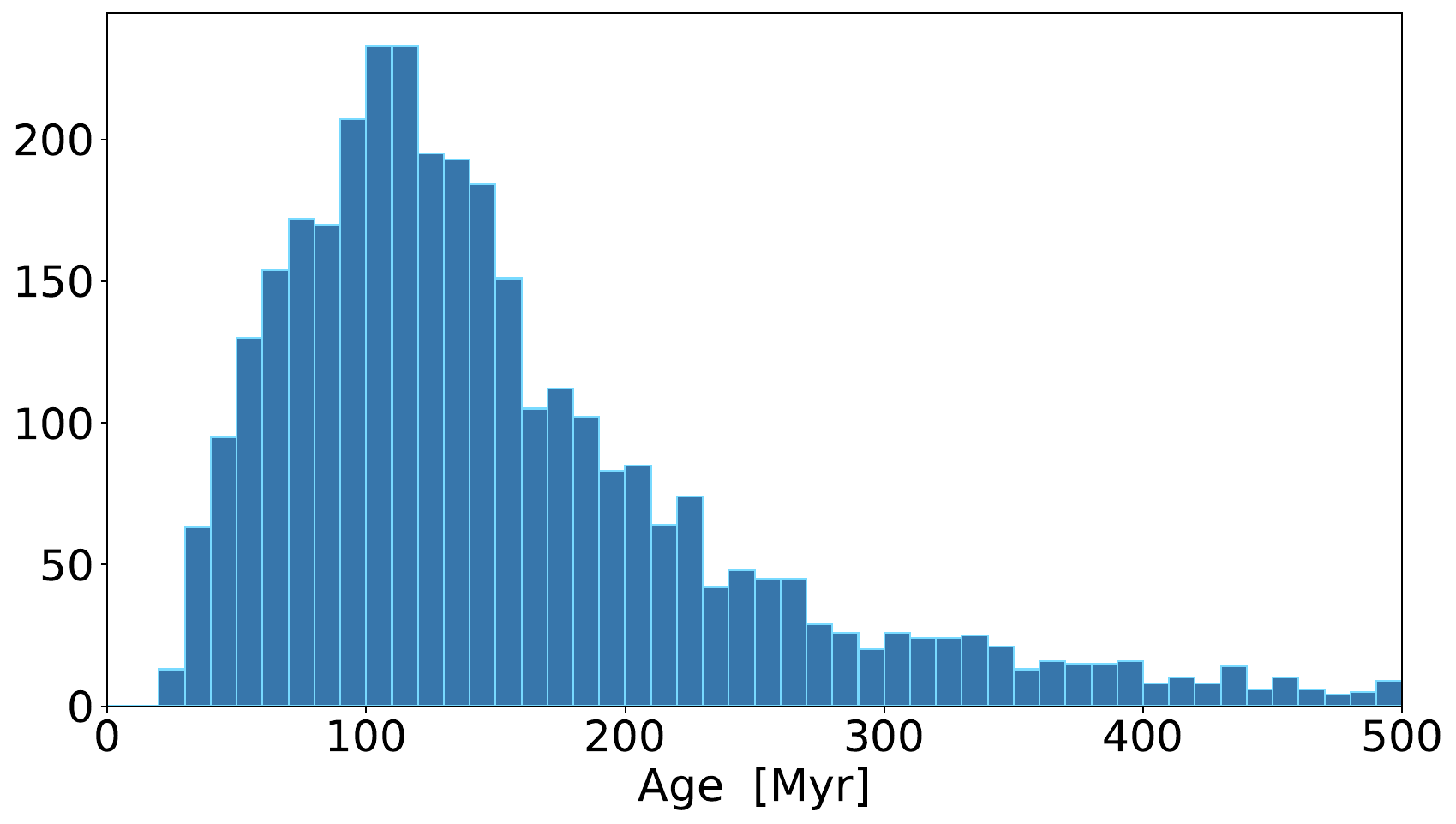}
    \caption{Ages of 3424 Cepheids in our sample estimated from the period-age relation derived by \citet{Anderson2016}. The median age is 132 Myrs.}
    \label{fig:ages}
\end{figure}

The period-age relation also depends on metallicity. Since spectroscopic metallicities are known only for a small subset of Galactic Cepheids, we estimate Cepheid metallicities with the metallicity gradient of \citet{Genovali2014}. The period-age relations are provided for three metallicity values (solar, LMC, and SMC). However, many Cepheids in the sample are more metal-rich, we therefore extrapolate the relation to cover the wider range of metallicity values. The calculated ages of our Cepheids vary between 24 Myr and 1.2 Gyr, a mean value of 163 Myr, and only 10\% are older than 300 Myr. Figure~\ref{fig:ages} shows the histogram of Cepheid ages.

\cite{Anderson2016} note that the estimated ages of Cepheids, when the crossing numbers, the position in the instability strip, and the rotational histories are not known, are uncertain to $\sim 50\%$. Consequently, these should not be used individually, but are sufficient for large sample studies.

\section{Discussion}
\label{discussion}

\begin{figure*}
    \centering
    \includegraphics[width=0.49\textwidth]{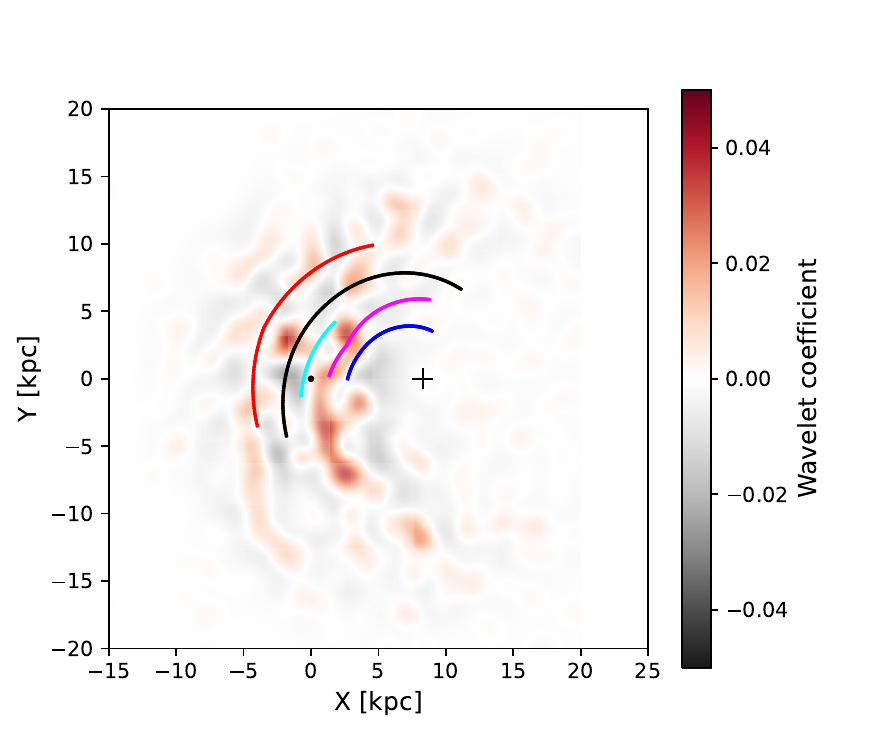}
    \includegraphics[width=0.49\textwidth]{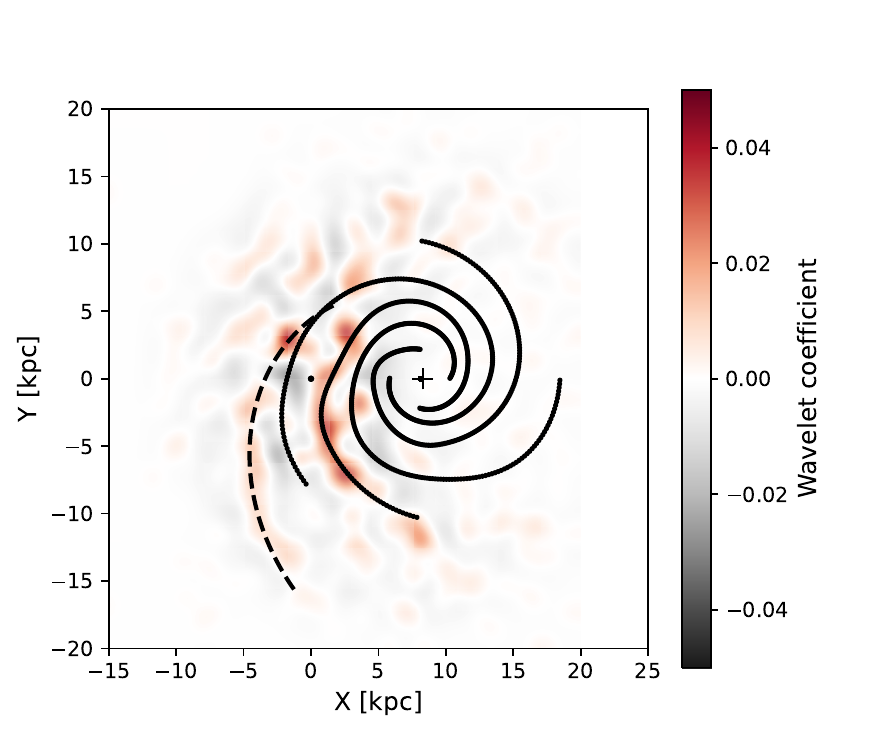}
    \caption{\emph{Left panel}: wavelet transformation at scale $j=4$ of the Cepheids sample, compared to the spiral arm model from \citet{Reid2019}. The black dot shows the Sun's position, while the black cross at X=8.277 kpc \citep[assuming the distance inferred by][]{GravityCollaboration:2022}, Y=0 kpc shows the Galactic center.  
    \emph{Right panel}: same as left panel, but now compared to the spiral arm model from \citet{Taylor:1993} (solid lines) and arm n.2 from \citet{Levine2006} (dashed line).}
    \label{fig:wavelet_comparison_models}
\end{figure*}

With our new distances it is worthwhile to at least make some brief remarks on the resulting spatial distribution of our Cepheids. 
Figure \ref{fig:wavelet_comparison_models} shows an overdensity map of the Cepheid sample in the Galactic disk, compared to different spiral arm models available in the literature, including that of  \citet{Reid2019} based on masers (left panel), and that of \citet{Taylor:1993} and \citet{Levine2006} based on the distribution of HII regions and neutral hydrogen (right panel).
The background overdensity map has been obtained with the 2-dimensional wavelet transformation (WT) code \emph{MGwave} by \cite{Lucchini:2023}, applying the \textit{à trous} algorithm \citep{Starck1994,Starck1998} to the X-Y histogram of our sample (bin size of 75~pc).  In particular, we show the wavelet scale $j=4$, which highlights mostly structures with sizes 
between 1.2 to 2.4 kpc.
We can distinguish at least two main features emerging from the (red) overdense regions, corresponding to segments of the Perseus arm (X $\sim$ -4 kpc, Y from $\sim$ -14 to 0 kpc) and the Sagittarius-Carina arm (X $\sim$ 2 kpc, Y from $\sim$ -8 to $\sim$ 1 kpc).

The comparison between the WT map and various spiral arm models available in the literature indicates that the geometry proposed by \cite{Levine2006} (black dashed line in Figure \ref{fig:wavelet_comparison_models}) based on HI data is the one that better traces the continuation of the Perseus arm in the third quadrant as traced by the Cepheids, as already remarked by previous works \citep{Poggio2021, GaiaCollaboration2023}. On the other hand, we do not necessarily expect to see a perfect agreement between the distribution of classical Cepheids and the models based on much younger tracers, such as the masers, as Cepheids have already had time to migrate from their birthplaces (the regions of highest star formation), if the angular rotation rate of stars is different than the spiral pattern speed \citep[][]{Skowron2019a}. Indeed, the coincidence of the Cepheids and the Levine arm seen in the HI would suggest that this arm is transient.
A more detailed analysis of the spiral structure using this dataset will be presented in Paper II \citep[][, in press]{Drimmel2024}. 

\begin{figure*}
    \centering
    \includegraphics[width=0.49\textwidth]{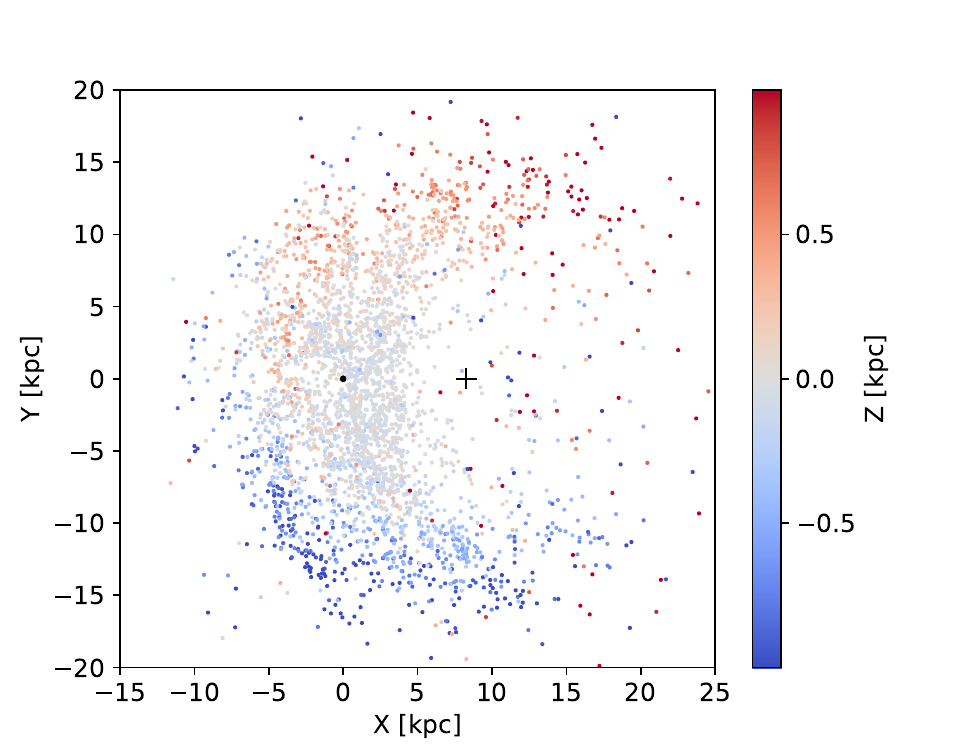}
    \includegraphics[width=0.49\textwidth]{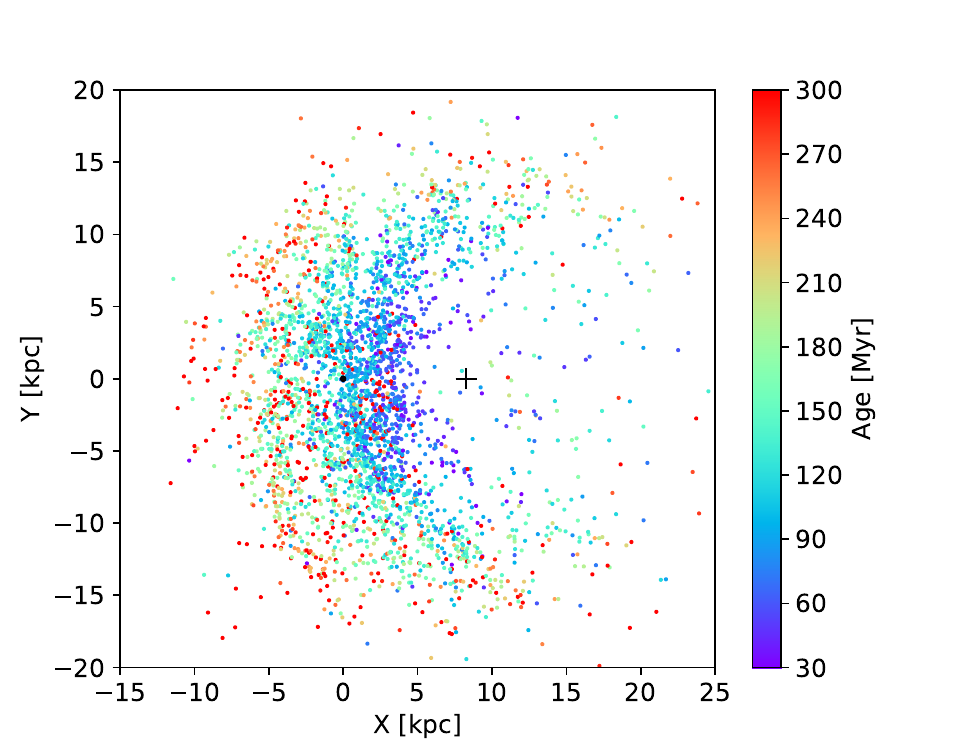}
    \caption{Position of the Cepheids in the Galactic plane. The Sun's position is shown by a black dot, and the Galactic center by a black cross. Stars are color-coded by their height above/below the Galactic plane Z (left panel) and age (right panel).}
    \label{fig:warp_age}
\end{figure*}

It is also worth noting that the vertical coordinates exhibit a clear signature of the Galactic warp, with the Cepheids being systematically shifted upwards (downwards) with respect to the Galactic plane for Y$\gtrsim$0 (Y$\lesssim$0), approximately up to $\sim$ 1 kpc (-1 kpc) in Z, as shown in Figure \ref{fig:warp_age} (left panel). The observed warp amplitude is in agreement with the prediction from warp models based on Galactic Cepheids available in the literature \citep[e.g.][]{Skowron2019a,Skowron2019b,Lemasle2022,Cabrera-Gadea2024,Poggio:2024}. 

The right panel of Figure~\ref{fig:warp_age} shows again the distribution of Cepheids in the Galactic plane, but now color-coded by age. As already remarked by previous works \citep[e.g.][]{Skowron2019a,Anders2024}, Cepheids in the outer disk are typically older than those in the inner parts due to the Galactic metallicity gradient, because less massive (i.e., shorter period, older) stars do not reach the instability strip during their evolution, when in high-metallicity environments. However, simulations by \cite{Skowron2019a} show that even after accounting for the metallicity gradient, we should observe more long-period Cepheids in the outer disk (farther than 16~kpc from the Galactic center), and less in regions with solar metallicity, assuming a constant star formation rate. This indicates, that such assumption might not be correct, and simulations show that in order to reproduce the current locations of Cepheids, star formation should happen in several distinct episodes in the past (for details see the supplementary material in \citealt{Skowron2019a}). 

In section \ref{sec:validation} we verified our uncertainties by comparing our photometric distances to parallax-based astrometric distances, which also resulted in an estimate of the mean parallax offset (zeropoint) for the subsample of 910 Cepheids used in the analysis. 
L21 characterized the parallax offset $Z_{\rm off}$ in \gdrthree\ as a function of color, magnitude and position on the sky, such that $\varpi - Z_{\rm off}$ is intended to be the true parallax (i.e. $Z_{\rm off} = -\varpi_0$ as defined in Eq.~\ref{eq:parallax}). If we recover their proposed offsets for our subsample of 910 Cepheids with good parallaxes, we find a mean offset $Z_{\rm off}$ of $-24.3 \: \muas$. Our estimate of $\varpi_0 \approx 7\: \muas$ suggests that the L21 offsets over-corrects the \gdrthree\ parallaxes for our sample. Meanwhile, the standard deviation of these L21 offsets is $14.7\: \muas$, while our measure of the dispersion of the zero point, $\sigma_0 \approx 27\: \muas$, is significantly larger than this.  However, our parameter $\sigma_0$ may also be capturing the additional uncertainty due to $\varpi_0$ varying on small angular scales \citep{Lindegren2021a} as well as absorbing possible under-estimation of the parallax uncertainties at bright magnitudes \citep{Sanders2023}.  Since the parallax offset has a complicated dependency on color, apparent magnitude and position on the sky, and is calibrated using quasars -- sources that are quite different from the Cepheids -- it is perhaps not surprising that our results differ from L21, who in any case encourages those who use the \gaia\ parallaxes to derive their own offsets appropriate for their science case.  Nevertheless it is useful to compare our derived offset with other studies using Cepheids. 

Most recently \citet{Molinaro:2023} re-derived period-Wesenheit relations for optical and NIR passbands from a selected subset of 443 Cepheids, including both metallicity and parallax offset terms.  In performing their calibration they use the \gdrthree\ parallaxes {\em after} applying the L21 correction, that is the true parallax is taken to be $\varpi_i - Z_{{\rm off},i} + \epsilon$, where $\epsilon$ is the {\em residual} parallax offset, and find $\epsilon = -22 \pm 4\: \muas$. Recovering the L21 offsets for their sample of stars we find a mean $Z_{\rm off}$ of $-26\: \muas$, implying a mean $\varpi_0$ of about $4\: \muas$. Their result is in good agreement with those of \citet{Bhardwaj:2021}, derived from a calibration of the PL relation for RR-Lyrae, while \citet{Cruz-Reyes2023} use a subset of 225 Cepheids, including some known open cluster members, and find a residual parallax offset of $-17 \pm 5\: \muas$. The mean L21 offset for their sample is $-24\: \muas$, implying a mean $\varpi_0$ of about $7\: \muas$, in good agreement with our estimate.  In any case, one should expect some variation in these estimates as they use different samples with differing distributions in apparent magnitude and color.

\section{Summary}

We have used both AllWISE and unWISE $W1$ photometry to estimate new distances for 3424 Galactic Classical Cepheids, which we make available in the form of a catalog. Using a subset of 910 Cepheids with good parallaxes we have confirmed that our distances are accurate with relative distance uncertainties on the order of 6\%. 

The accuracy of our $W1$-based distances is in part due to the relatively small extinction correction needed in the mid-IR. Indeed, for 90\% of our sample the $A_{W1}$ extinction is estimated to be less than 0.45 mag, with a mean value of 0.22 mag. In addition, we correct for extinction in the mid-IR using a 3D extinction map, rather than using a Wesenheit index, to avoid systematic errors resulting from assuming a universal extinction curve. In the optical this can lead to large systematic errors in final distances, while in the mid-IR imprecise colors will increase the photometric uncertainties of Wesenheit magnitudes.
We instead correct for extinction using a 3D extinction map in the $K_S$ band and adopting an extinction ratio of $A_{W1}/A_{K_S} = 0.47$ derived from the mid-IR extinction curve from \cite{Gordon2023}. While our adopted value for $A_{W1}/A_{K_S}$ assumes a ratio of total-to-selective absorption $R_V = 3.1$, this extinction ratio varies by less than 10\% from this mean value for $R_V$ values between 2.6 and 3.6, and would result in a relative distance uncertainty of not more than 2.1\%, typically around $1\%$.
In contrast, the variation of $R_V$ in the same range of 2.6 to 3.6 would vary the Wesenheit magnitude $W_G$ by $\pm 0.6$ mag (for the typical Cepheid in our sample), resulting in a relative distance uncertainty of 30\%.

In addition, the precision of our distances is also thanks to the tight correlation of the PL diagram in the $W1$ passband, so that the scatter about the PL relation only contributes about 0.08 mag to the uncertainty in the luminosities \citep{Wang2018}. 

In deriving our distances we rely on the published PL relation of \citet{Wang2018} whose calibration is based on \gdrtwo\ parallaxes. It would be beneficial in the near future to redo this calibration using the \gdrthree\ parallaxes including the parallax offset of the sample in the adjustment of parameters.  The need for a metallicity term still remains an open question in the mid-IR, though the addition of such a term can potentially introduce systematic errors if the metallicities used are not on the same metallicity scale as those used to calibrate such a term. Traditionally metallicity calibrations have relied on a subset of bright Cepheids with high-resolution spectroscopy, with the fainter Cepheids using a metallicity inferred from an adopted metallicity gradient. However such a gradient will necessarily also have an intrinsic scatter which will be an additional source of uncertainty that may well cancel any potential benefit from adopting a metallicity term. 

Galactic Cepheids have played an important role in understanding the structure and kinematics of the young stellar disk. 
In this contribution we have only briefly commented on the distribution of our sample. In an accompanying paper we will explore in more detail what this sample reveals about the spiral structure in our Galaxy \citep[][, in press]{Drimmel2024}.  We hope that additional future Galactic studies will benefit from using these distances. 

\section*{Acknowledgments}

We thank the Referee for useful comments that improved the manuscript, and to Robert Benjamin for providing a compilation of the \textit{Glimpse} surveys. DMS thanks Paweł Pietrukowicz for fruitful discussions on classical Cepheid classification.
RD thanks Rene Andrae and Federico Marocco for useful conversations, Morgan Fouesneau for computing extinction ratios, and Vincenzo Ripepi and Roberto Molinaro for providing the list of sources used in 
\citet{Molinaro:2023}. We acknowledge discussions with Edward Schlafly on handling unWISE uncertainties.

DMS acknowledges support from the European Union (ERC, LSP-MIST, 101040160). Views and opinions expressed are however those of the authors only and do not necessarily reflect those of the European Union or the European Research Council. Neither the European Union nor the granting authority can be held responsible for them.

RD \& EP are supported in part by the Italian Space Agency (ASI) through contract 2018-24-HH.0 and its addendum 2018-24-HH.1-2022 to the National Institute for Astrophysics (INAF).

SK \& RD acknowledge support from the European Union's Horizon 2020 research and innovation program under the GaiaUnlimited project (grant agreement No 101004110).

SK acknowledges use of the INAF PLEIADI@IRA computing resources.

This work has made use of data from the European Space Agency (ESA) mission
{\it Gaia} (\url{https://www.cosmos.esa.int/gaia}), processed by the {\it Gaia}
Data Processing and Analysis Consortium (DPAC,
\url{https://www.cosmos.esa.int/web/gaia/dpac/consortium}). Funding for the DPAC
has been provided by national institutions, in particular the institutions
participating in the {\it Gaia} Multilateral Agreement.

This publication makes use of AllWISE data products derived from the Wide-field Infrared Survey Explorer, which is a joint project of the University of California, Los Angeles, and the Jet Propulsion Laboratory/California Institute of Technology, and NEOWISE, which is a project of the Jet Propulsion Laboratory/California Institute of Technology. WISE and NEOWISE are funded by the National Aeronautics and Space Administration.

This project was developed in part at the Lorentz Center workshop "Mapping the Milky Way", held 6-10 February, 2023 in Leiden, Netherlands.

\software{
    TOPCAT 
    \citep{2005ASPC..347...29T,2006ASPC..351..666T},
    Julia 
    \citep{Julia2012},
    Matplotlib \citep{Hunter:2007},
    IPython \citep{PER-GRA:2007}, 
    SciPy \citep{2020SciPy-NMeth},
    Astropy \citep{astropy:2013,astropy:2018,astropy:2022},
    NumPy \citep{harris2020array}
    } 
    
\appendix

\section{Photometric accuracy of unWISE blended sources }
\label{Appendix:A}

In order to verify the photometric accuracy of the blended sources, we derived a photometric transformation from the \gaia\ photometric system to $W1_U$ and checked if the residuals show a dependence on the parameter {\tt fracflux} that represents the PSF-weighted fraction of flux from the source.

We identified a crowded region close to the Galactic plane, $205\deg < l < 215\deg$ and $15\deg<b<16\deg$, with low extinction E(B-V) $< 0.1$, where we selected single sources cross-matched between \gaia~DR3 and unWISE. We also included robust data-driven metallicities [M/H] derived from \gaia\ XP spectra by \citet{Andrae2023}.
In total, we selected 8723 dwarfs that satisfy the following criterion:
\medskip

{\tt
w1mpro\_unwise > 8.5 AND w1mpro\_unwise < 15 AND phot\_g\_mean\_mag < 16 AND number\_of\_mates = 0 AND classprob\_dsc\_combmod\_star > 0.95 AND angular\_distance < 0.2 AND ffw1\_unwise > 0.99 AND ffw1\_unwise <1 AND 
logg\_gspphot > 4.0 AND mh\_xgboost > -1  AND mh\_xgboost < +0.4}
\medskip

\begin{figure}
    \centering
    \includegraphics[width=0.49\textwidth]{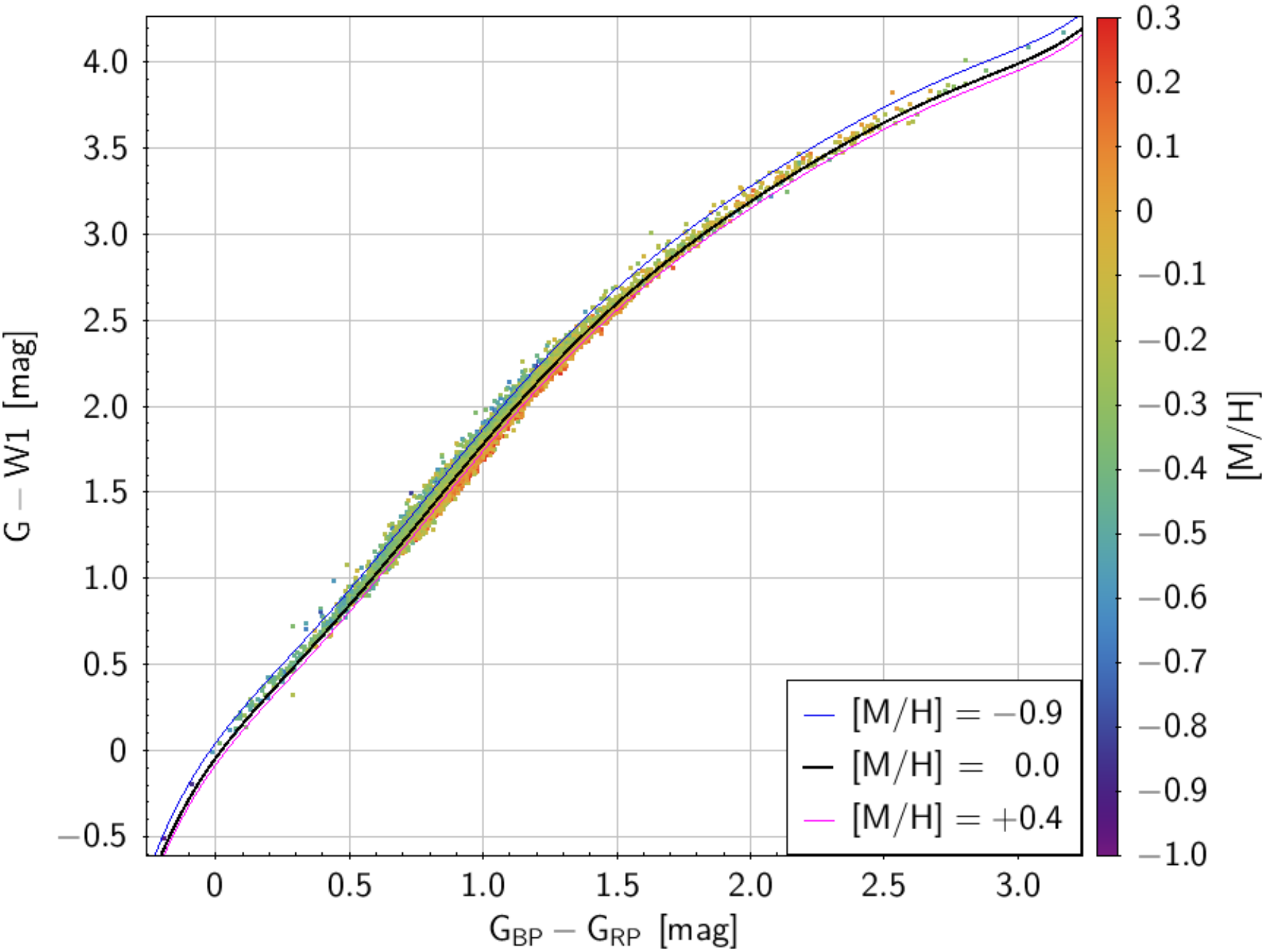}
    \caption{color-color distribution $G-W1$ vs.\ $G_{\rm BP}-G_{\rm RP}$ of 8723 dwarfs color coded by metallicity.  Lines with different colors represent the photometric transformation of Eq.\ \ref{eq:colourtransform} for [M/H]=0 (black), [M/H]$=+0.4$ (violet), and [M/H]$=-0.9$ (blue).}
    \label{fig:color_transformation}
\end{figure}

We adopted a photometric transformation based on a 7$^{\rm th}$ order polynomial function of the \gaia\ color $(G_{\rm BP}- G_{\rm RP})$, plus an additional metallicity term, as follows:

\begin{equation}
    G - W1 = \sum_{n=0}^7  p_n \cdot (G_{\rm BP} - G_{\rm RP} )^n + p_8 \cdot [{\rm M/H}]
    \label{eq:colourtransform}
\end{equation}

where [M/H] is the metallicity estimated by \citet{Andrae2023}.  As shown in Figure \ref{fig:color_transformation}, Equation \ref{eq:colourtransform} provides an accurate transformation for dwarfs in a wide color range, $ -0.2 < (G_{\rm BP} - G_{\rm RP} ) < 3$. 
The standard deviation of the best fit residuals is 0.032 mag and the estimated parameters are listed in Table \ref{tab:parameters}.

Finally we applied the transformation \ref{eq:colourtransform} to a larger test sample of 160,806 dwarfs, including blended sources with values of {\tt fracflux} down to $\sim 0.2$, selected as follows:\\
{\tt
w1mpro\_unwise > 8.5 AND w1mpro\_unwise < 15 AND phot\_g\_mean\_mag < 16 AND number\_of\_mates = 0 AND classprob\_dsc\_combmod\_star > 0.5 AND angular\_distance < 2 AND ffw1\_unwise <1  AND logg\_gspphot > 4.0 AND b > 12 AND b < 20 AND mh\_xgboost > -1.0  AND mh\_xgboost < 0.6
}\\

\begin{figure}
    \centering
    \includegraphics[width=0.49\textwidth]{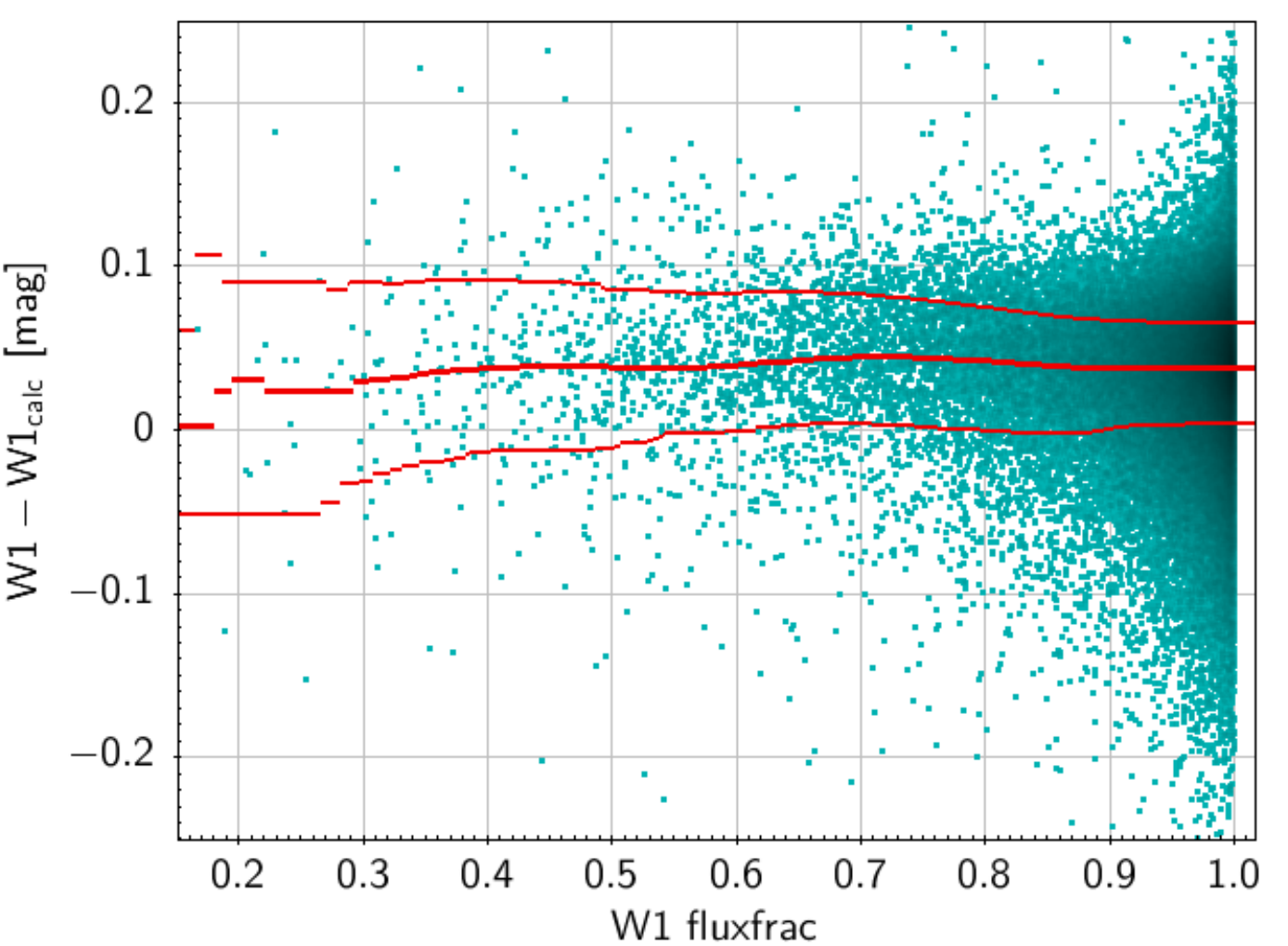}
    \caption{Test sample: distribution of the residuals between $W1_U$ and the value computed from Eq.\ \ref{eq:colourtransform} vs.\  {\tt fracflux}. Red lines evidence the median, the 16th and 84th percentiles.}
    \label{fig:deltaW1_vs_ffW1}
\end{figure}

Figure \ref{fig:deltaW1_vs_ffW1} shows the distribution of the residuals between the unWISE magnitude $W1_U$ and the value computed from Eq.\ \ref{eq:colourtransform}.
We notice a mild decrease of the dispersion as a function of {\tt fracflux}.

We concluded that  blending does not significantly affect the photometric accuracy of our tracers having  {\tt fracflux} $> 0.5$ for more than 95\% of the sources.

\begin{table}
\caption{Parameters of the photometric transformation}
\label{table:parameters}
\centering
\begin{tabular}{c r r }
\hline\hline
Parameter & value & standard error \\
\hline
    $p_0$ & $-0.04450$ & 0.00759  \\
    $p_1$ & 2.13784  &   0.05459     \\
    $p_2$ & $-2.06468$ &   0.18878  \\
    $p_3$ &  4.22223  &  0.32283 \\
    $p_4$ & $-3.83099$ &    0.29329\\
    $p_5$ & 1.69811  &  0.14390 \\
    $p_6$ & $-0.36988$ &  0.03585 \\
    $p_7$ & 0.03179 &  0.00354 \\
    $p_8$ & $-0.09839$ &  0.00164\\ 
    
\hline
\end{tabular}
\label{tab:parameters}
\end{table}

\bibliography{paper}{}
\bibliographystyle{aasjournal}

\end{document}